\documentclass[10pt,journal,compsoc]{IEEEtran}

\ifCLASSOPTIONcompsoc
  \usepackage[nocompress]{cite}
\else
  \usepackage{cite}
\fi

\usepackage{hyperref}
\usepackage{wrapfig}
\usepackage{color}
\usepackage{xcolor}
\usepackage{amsmath}
\usepackage{placeins}
\usepackage[abs]{overpic}
\usepackage{tabularx, colortbl}
\usepackage[ruled, vlined]{algorithm2e} 
\usepackage{float}
\usepackage{bm}

\newcommand{\para}[1]{\noindent\textbf{#1}.}

\begin{document}

\title{SEG-MAT: 3D Shape Segmentation Using Medial Axis Transform}

\author{Cheng~Lin,~
        Lingjie~Liu,~
        Changjian~Li,~
        Leif~Kobbelt,~
        Bin~Wang,~
        Shiqing~Xin,~
        Wenping~Wang  

\IEEEcompsocitemizethanks{
\IEEEcompsocthanksitem C. Lin, L. Liu, C. Li and W. Wang are with The University of Hong Kong. E-mail: chlin@hku.hk, liulingjie0206@gmail.com, chjili2011@gmail.com, wenping@cs.hku.hk
\IEEEcompsocthanksitem L. Kobbelt is with RWTH Aachen University. 
E-mail: kobbelt@cs.rwth-aachen.de
\IEEEcompsocthanksitem B. Wang is with Tsinghua University and Beijing National Research Center for Information Science and Technology (BNRist). 
E-mail: wangbins@tsinghua.edu.cn
\IEEEcompsocthanksitem S. Xin is with Shandong University. 
E-mail: xinshiqing@gmail.com

}
\thanks
}

\IEEEtitleabstractindextext{%
\begin{abstract}
Segmenting arbitrary 3D objects into constituent parts that are structurally meaningful is a fundamental problem encountered in a wide range of computer graphics applications. Existing methods for 3D shape segmentation suffer from complex geometry processing and heavy computation caused by using low-level features and fragmented segmentation results due to the lack of global consideration. We present an efficient method, called {\em SEG-MAT}, based on the medial axis transform (MAT) of the input shape. Specifically, with the rich geometrical and structural information encoded in the MAT, we are able to develop a simple and principled approach to effectively identify the various types of junctions between different parts of a 3D shape. Extensive evaluations and comparisons show that our method outperforms the state-of-the-art methods in terms of segmentation quality and is also one order of magnitude faster. 
\end{abstract}

}

\maketitle

\IEEEdisplaynontitleabstractindextext

%
\IEEEpeerreviewmaketitle

\IEEEraisesectionheading{\section{Introduction}}
\label{sec:introduction}

\begin{figure}[t]
\vspace{-4mm}
    \begin{overpic}[width=\linewidth]{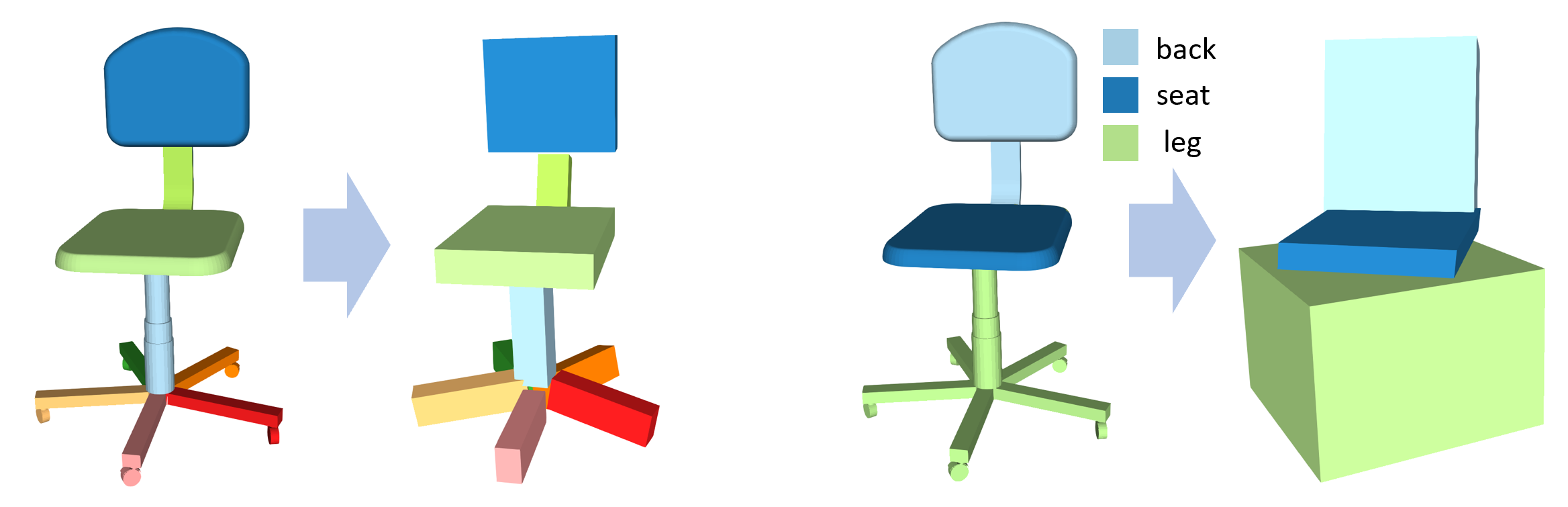}
    \put(37, -5) {\small SEG-MAT}
    \put(18,-13) {\small (geometrical analysis)}
    \put(162, -5){\small ShapePFCN \cite{kalogerakis20173d}}
        \put(153,-13) {\small (semantic learning)}
    \end{overpic}
    \vspace{0mm}
    \caption{Difference between geometrical and semantic segmentation methods. The geometrical methods segment a shape into part instances and usually produce weakly convex components; the semantic learning methods focus on the correspondences between the pre-defined labels and shapes without considering detailed part instances. }
    \label{fig:two-cat-diff}
\end{figure}
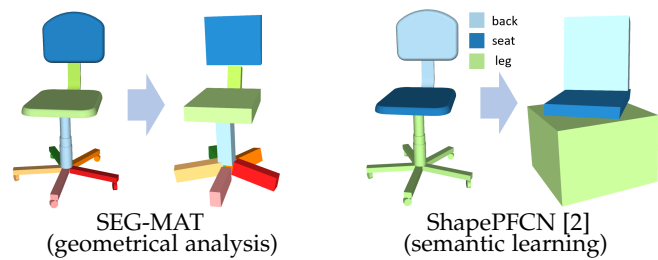

\IEEEPARstart{A}utomatically segmenting 3D shapes into structurally simple parts is an important problem in many applications of computer graphics and computer vision. Most existing methods can be roughly categorized into two classes based on their goals and techniques. The first class of methods is based on supervised learning and relies on annotated semantic labels. Basically, the semantic meanings are pre-defined consistently within a shape category, and then the semantic labels are manually annotated on datasets to train the neural networks. The second class of methods uses geometrical analysis and can be either rule-based or unsupervised learning-based. These methods decompose a shape by finding part boundaries where certain geometrical properties are met. Semantic \footnote{The``semantic'' and ``geometrical'' methods refer to the techniques used, i.e., learning from semantic labels or analyzing geometrical properties.} and geometrical segmentation methods have different criteria and goals. The semantic segmentation methods aim to find the correspondences between shapes and pre-defined labels, but do not focus on the part instances; the geometrical segmentation methods predominantly follow the minima rule \cite{hoffman1984parts}, that is, each part should be weakly convex, which is shown in Fig. \ref{fig:two-cat-diff}. These methods also have different technical merits and applications in different problem settings. The semantic methods are used to repeatedly identify semantic meanings for a group of similar shapes in the same category, while handling a unique shape or an unseen category is out of its scope; the geometrical methods are not limited to pre-defined labels or categories, but they cannot give the interpretable semantic labels of the parts. 

\begin{figure}[t]
    \begin{overpic}[width=\linewidth]{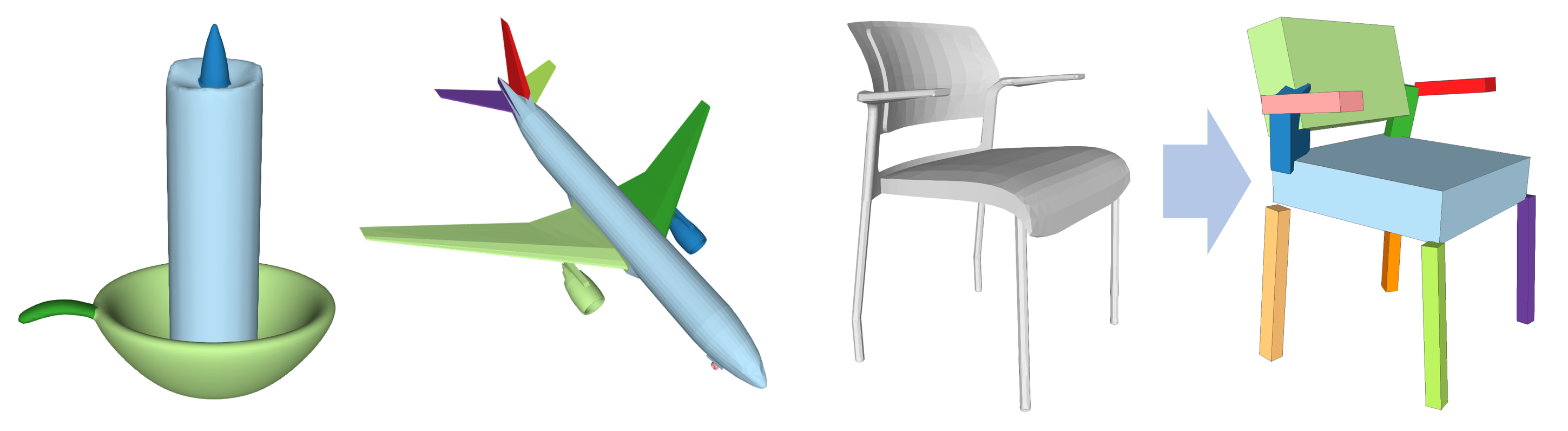}
    \end{overpic}
    \vspace{-6mm}
    \caption{Our algorithm is a geometry-driven method for single shape segmentation and analysis.}
    \label{fig:teaser}
\end{figure}

In this paper, we return to the classic method based on the geometrical analysis for segmenting an arbitrary 3D shape (see Fig.~\ref{fig:teaser}). 3D shape segmentation driven by geometrical analysis can be used to guide various tasks including modeling \cite{modelingbyexample2004,lin2020modeling}, retrieval \cite{zuckerberger2002polyhedral,shaperetrieval2010}, mesh multi-resolution and compression \cite{multires_compression2011}, texture mapping \cite{levy2002texturemapping}, reverse engineering \cite{reverseengineering2007}, etc. Also, a robust geometry-based shape segmentation method can facilitate the data annotation process of various data-driven methods for 3D shape analysis.

Although shape segmentation by geometrical analysis is a well-researched problem, existing methods still have some notable issues. (i) A desirable part may have complex geometry and varying properties. Therefore, the methods using a single feature descriptor struggle to characterize the geometry of a part \cite{sdf2008}, while extracting various features to encode low-level geometrical properties needs complex optimizations and a large quantity of computation time \cite{shu2016unsupervised, wcseg}. (ii) The extracted local geometrical features lack high-level structure information of shape context, which tends to give counter-intuitive segmentation results. (iii) Many approaches need to tune shape-specific parameters, such as indicating the number of parts to perform a clustering algorithm for different shapes or categories \cite{randcut}, which creates a burden on users.

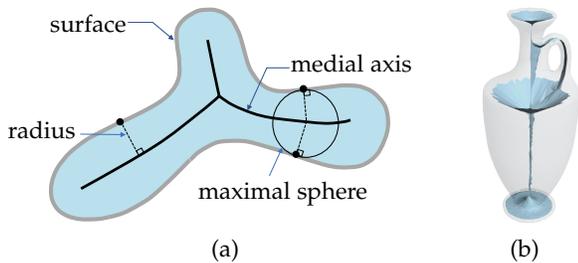
\begin{figure}[!htb]
\vspace{-4mm}
\hspace{3.5mm}
    \begin{overpic}[width=\linewidth]{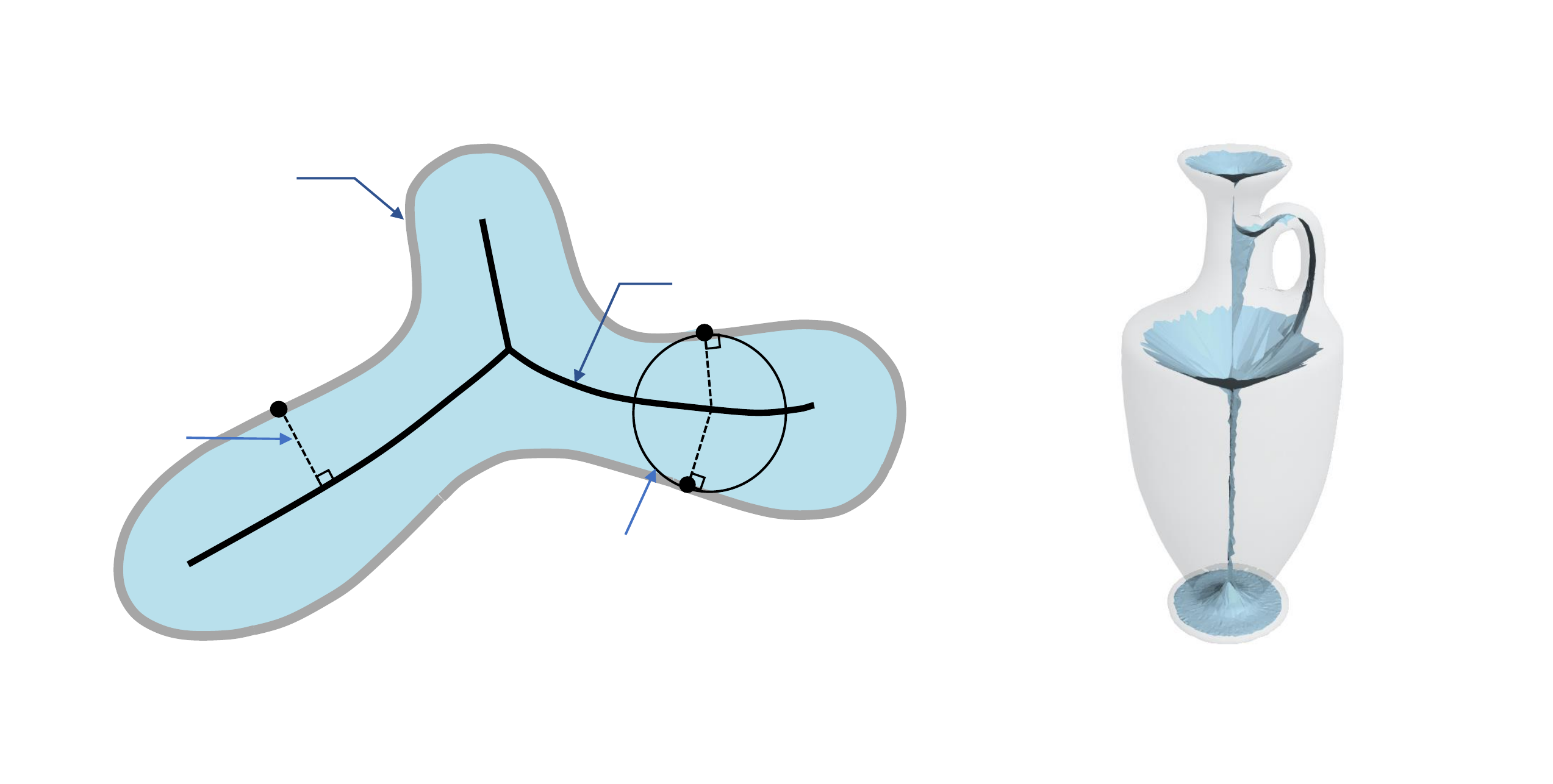}
        \put(80, 10) {\small (a)}
        \put(193, 10) {\small (b)}
        \put(3,55){\small radius}
        \put(19,95){\small surface}
        \put(110,80){\small medial axis}
        \put(75,32){\small maximal sphere}
    \end{overpic}
    \vspace{-8mm}
    \caption{Illustration of the medial axis transform (MAT).  (a) The MAT of a 2D shape consists of a set of centers of the maximal inscribed circles together with the circle radii. (b) In 3D, the MAT consists of a set of centers of the maximal inscribed spheres of a shape together with the sphere radii.}
    \label{fig:matintro}
\end{figure}

We propose a new method based on the medial axis transform (MAT) \cite{blum1967transformation} for 3D shape segmentation.  An illustration of the MAT of a 2D and a 3D shape is shown in Fig.~\ref{fig:matintro}. Our key observation is that the MAT of a 3D shape encodes rich structural and geometrical information, which provides valuable clues for robust 3D shape segmentation.

Instead of extracting low-level features, we propose to use the MAT to analyze high-level junctions where different parts of a 3D shape meet. As a result, our method overcomes the aforementioned limitations of existing methods. First, the MAT is already a compact shape representation in which a series of geometrical clues are aggregated, such as thickness, angles and branches; therefore, it can characterize the geometrical variation jointly from different perspectives, which overcomes the issue (i).  Second, the MAT is globally informative which is reflected in two aspects: (1) it extracts the global structure of a shape; (2) since the MAT is a volume-based representation, segmenting the MAT is to handle volumetrical parts rather than local surface points. Thus, the global information is significantly reinforced in the shape analysis, which overcomes the issue (ii). Third, our method focuses on finding the high-level junctions instead of clustering low-level feature descriptors, enabling a simple but effective formulation without complex optimization and heavy computation; this overcomes the issues (i) and (iii). We will present experimental results later in the paper to demonstrate the improvements over the other existing methods. 

To summarize, our proposed method has the following strengths and contributions:

\begin{itemize}
    \item Our method is effective and simple. It outperforms the state-of-the-art methods on public datasets and is about one order of magnitude faster. Our method is also flexible and robust; it can generate reasonable segmentation results for different datasets and categories using the same set of parameters.
    
    \item We propose a novel formulation on the MAT for shape analysis. We make full use of its advantages to exploit the structural and geometrical information encoded, while circumventing its drawback, i.e., the notorious instability of MAT.
    
\end{itemize}

\section{Related Work}
\subsection{3D shape segmentation}
In the past decades, numerous works have been developed for 3D shape segmentation. The different categories of methods are discussed in detail below. Readers can also refer to \cite{rodrigues2018part} for a comprehensive survey.

\para{Supervised learning by semantic labels} This kind of methods \cite{kalogerakis2010learning, guo20153d, kalogerakis20173d, xu2017directionally} is learning-based, which aims to label an input shape with a set of pre-defined semantic meanings. With strong supervision, these methods are able to provide superior results with useful semantic labels. However, the limitations of these methods lie in the requirement of a large amount of manually labeled segmentation results. Also, the trained neural network can only be used to segment the shapes in the same category with the same semantic components, which limits their applicability to some extent. Note that methods of this kind have different goals, applications and technical merits with our method; therefore, we do not directly compare our method with them but give insightful discussions in Sec.~\ref{sec:semantic_discussion}.

\para{Rule-based segmentation by geometrical analysis}
Our method falls into this class. These methods aim to find the geometrical boundaries between parts where some geometrical properties are met, without considering explicit semantics labels or consistency across shapes. These methods usually extract certain feature descriptors on the surface of a shape for segmentation, such as weakly convex decomposition \cite{wcseg, asafi2013weak}, concavity-aware fields \cite{concaveawarefields}, randomized cuts \cite{randcut}, random walks \cite{randomwalk}, core extraction \cite{coreextra} and K-Means \cite{kmeans}. The development of these methods is predominantly guided by the minima rule \cite{hoffman1984parts}, which indicates that a complex shape is a composition of approximately convex parts and the cut boundaries lie along the concavities. These methods are normally validated and compared using the Princeton Segmentation Benchmark (PSB) \cite{chen2009benchmark}. The evaluations on the PSB show that our method outperforms the state-of-the-art methods and is about one order of magnitude faster.

\para{Unsupervised learning by geometrical analysis} In recent years, there has been a growing interest in unsupervised learning for geometrical segmentation. This kind of methods also targets discovering a shape's intrinsic geometry instead of pre-defined labels, but they resort to learning from large amounts of data. Shu \emph{et al.} \cite{shu2016unsupervised} propose to transform a set of handcrafted features using an auto-encoder and then segment the shape into different parts without any pre-defined labels. Tulsiani \emph{et al.} \cite{abstractionTulsiani17} propose an unsupervised method to generate a primitive-based representation to approximate a target 3D shape. Each part is represented by a cuboid and the network is trained using a reconstruction loss. Sun \emph{et al.} \cite{sun2019hierarchy} present an adaptive hierarchical cuboid representation for 3D shape abstraction. These methods all need to be trained separately on each shape category. In addition, cuboids cannot represent complex geometry, causing the difficulty of handling complex structures and shape details. Our method, however, is able to handle arbitrary 3D shapes using consistent settings without the need for training data. We compare our method with them and show better performance in terms of segmentation quality and flexibility (see Sec.~\ref{sec:cmp-learning}).

\para{Co-segmentation methods} Co-segmentation is a specific instance of shape segmentation problem. These methods \cite{coseg2012, coseg2013, shu2016unsupervised} take as input a collection of shapes that have some common characteristics and output consistent segmentation results across shapes. More recently, Chen \emph{et al.} introduce BAE-NET \cite{chen2019bae_net}, a branched autoencoder network for shape co-segmentation. These methods ensure the consistency of the segmentation across shapes within the same category, but cannot handle a single shape.

\subsection{Interior information for structure analysis}

\para{Skeletal representations} The closely relevant methods to SEG-MAT are those based on curve skeletons \cite{brunner2004mesh, reniers2008part, tierny2007topology}. These methods exploit the simplicity and component-wise structure of the shape skeletons to produce reasonable segmentation results. However, the curve skeletons are only empirically understood for tube-like parts but not mathematically defined for arbitrary shapes. Therefore, only articulated shapes can be properly represented by curve skeletons, and these methods can only be applied to a special class of shapes. In contrast, SEG-MAT overcomes this limitation by using the MAT which is rigorously defined as a complete representation for an arbitrary 3D shape. Feng \emph{et al.} \cite{feng2015skelcut} analyze the cut distribution based on the medial geodesic function \cite{dey2006defining} (MGF) computed with the medial surfaces to identify different parts for shape segmentation. The method solely relies on the single descriptor MGF without considering the shape structure; while SEG-MAT leverages various properties of the MAT, leading to higher-quality and more structure-aware segmentation results.

\para{Volumetric representations}  There are a few methods that use the volumetric information for shape segmentation, such as bounding volume computation \cite{lu2007variational}, part-aware metric \cite{liu2009part} and lines-of-sight \cite{asafi2013weak}. Zhou \emph{et al.} propose a quantitative measure of cylindricity for a shape part, enabling a shape decomposition method using generalized cylinders (GCD). In nature, the GCD is still based on the simplicity of curve skeletons so it inevitably exhibits the common limitations when applied to some non-tubular shapes. Shapira \emph{et al.} \cite{sdf2008} introduce shape diameter function (SDF) for 3D shape segmentation.  The SDF is used as an alternative to MAT for estimating the local thickness of an object, since existing techniques cannot handle the instability of MAT. In contrast, we directly use the MAT since our novel and simple formulations can address this problem. Our method also significantly outperforms the SDF method since the MAT is more structurally informative.

\subsection{Region growing for 3D segmentation}
The region growing technique is widely used for segmentation due to its simplicity. The super-face algorithm \cite{kalvin1996superfaces} uses a region growing strategy along with a set of representative planes for polygonal mesh simplification. Chazelle \emph{et al.}~\cite{chazelle1997strategies} present a method for convex decomposition that also uses a region growing with randomly selected starting faces. The watershed algorithm, which is effective for image segmentation, is in fact a region growing approach with multiple seed points. Accordingly, using the watershed algorithm for the segmentation of 3D surface mesh has also been studied  \cite{sun2002triangle,zuckerberger2002polyhedral,koschan2003perception}.

However, region growing is a greedy algorithm, and these methods locally focus on the faces of the surface mesh. Consequently, they all suffer from the over-segmentation problem and always output fragmented surface patches. In this paper, although the simple region growing strategy is used, it is performed on the MAT domain with novel energy terms and strategies, which results in superior segmentation results that are more in line with human cognition.

\section{Preliminary}

\subsection{Medial axis transform (MAT)}
The medial axis transform (MAT) \cite{blum1967transformation} of a 3D shape $S$ is defined by two parts: (1) the centers of a set of maximal spheres inscribed to the shape $S$, called medial axis; and (2) the radius function associated with each such sphere center. See Fig.~\ref{fig:matintro} for an illustration. In general, the MAT of a 3D shape $S$ consists of 2D non-manifold surface sheets and 1-D curve segments, with the radius function defined on this complex. The MAT is a complete shape descriptor, which means that it can be used to reconstruct the original domain. By its definition, the MAT explicitly encodes both structural and geometrical natures in the interior of a 3D shape, which is highly informative for shape analysis. 

\subsection{MAT graph}
\label{sec:matdef}

\begin{figure}[!htb]
    \begin{overpic}[width=\linewidth]{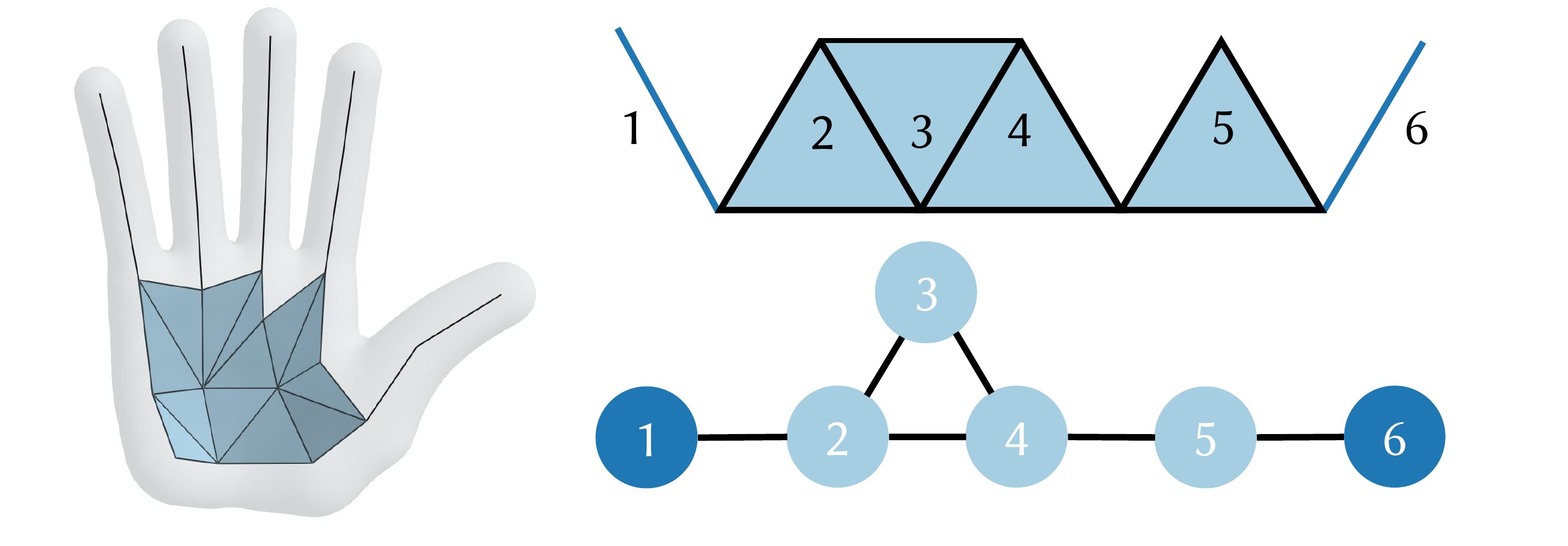}
        \put(40, -5) {\small (a)}
        \put(160, -5) {\small (b)}
    \end{overpic}
        \vspace{-5mm}
    \caption{Visualization of the medial mesh and the definition of the MAT graph. (a) The medial mesh of a hand model. (b) The MAT graph representation with the face nodes and the edge nodes.}
    \label{fig:matgraph}
\end{figure}

\begin{figure}[!htb]
    \centering
    \begin{overpic}[width=0.85\linewidth]{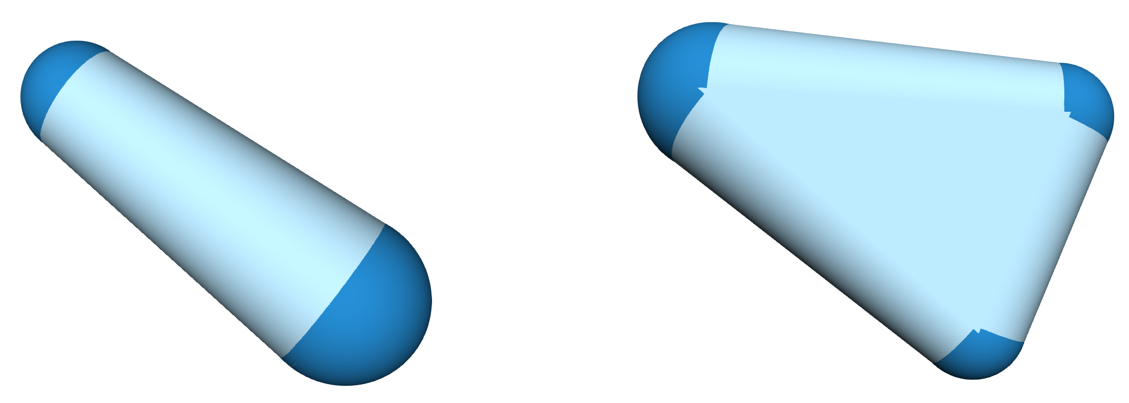}
        \put(40, -5) {\small (a)}
        \put(160, -5) {\small (b)}
    \end{overpic}
    \caption{The medial primitives. We show a cone (a) and a slab (b) given by the interpolation of their medial spheres. }
    \label{fig:slab}
    \vspace{-5mm}
\end{figure}

\para{Medial mesh} The key to SEG-MAT is an MAT represented as a triangle mesh, called a {\em medial mesh} (see Fig.~\ref{fig:matgraph} (a)). A vertex $m_i = (c_i,r_i)$ of the mesh denotes the maximally inscribed sphere centered at the point $c_i$ with radius $r_i$. Let $e_{ij}=\{m_i;m_j\}$ denote the edge of the mesh connecting two vertices $m_i$ and $m_j$. Then the edge $e_{ij}$ defines a {\em medial cone}, which is the convex hull of the two spheres $m_i$ and $m_j$ (see Fig.~\ref{fig:slab} (a)).  Let $f_{ijk}=\{m_i;m_j;m_k\}$ denote the triangle face with the vertices $m_i$, $m_j$, and $m_k$. Similarly, the face $f_{ijk}$ defines a {\em medial slab}, which is the convex hull of the three spheres $m_i$,  $m_j$, and $m_k$ (see Fig.~\ref{fig:slab} (b)). The input 3D shape is then approximated by the union of all these medial cones and medial slabs. 

\para{MAT graph} Medial cones and medial slabs serve as the elementary primitives in our segmentation operations. Therefore, we need to encode the connectivity between these primitives by defining the {\em MAT graph} on the medial mesh, as shown in Fig. \ref{fig:matgraph} (b). There are two types of nodes in the MAT graph: (1) a face node $N^f$ given by a triangle face $f=\{m_i;m_j;m_k\}$; (2) an edge node $N^e$ given by an edge $e=\{m_i;m_j\}$. We use $\overline{R}(N_i)$ to denote the average medial sphere radius of a face node or an edge node $N_i$. Two nodes of the MAT graph are connected if they share at least one vertex on the medial mesh.

\para{MAT computation} The MAT of a 3D object is known to be sensitive to boundary noise. Over the years some effective methods \cite{li2015q, yan2018voxel} have been proposed for producing a compact and accurate MAT of a 3D object. In this paper, we use the Q-MAT \cite{li2015q} method to generate the input MAT to our method.

\subsection{Segmentation criteria}
\label{sec:junctions}
We use the term {\em junction} to refer to a common boundary along which adjacent parts or segments meet. The mathematical model that characterizes these junctions is based on a number of segmentation criteria. Instead of proposing (a combination of) low-level feature descriptors, we capture higher-level shape context by expressing the segmentation criteria in terms of MAT properties. Inspired by well-established Gestalt principles \cite{todorovic2008gestalt}, we define a hierarchical taxonomy of junction types based on the topological and geometrical properties of the MAT.

\begin{figure}[!htb]
\centering
\vspace{-2mm}
    \begin{overpic}[width=\linewidth]{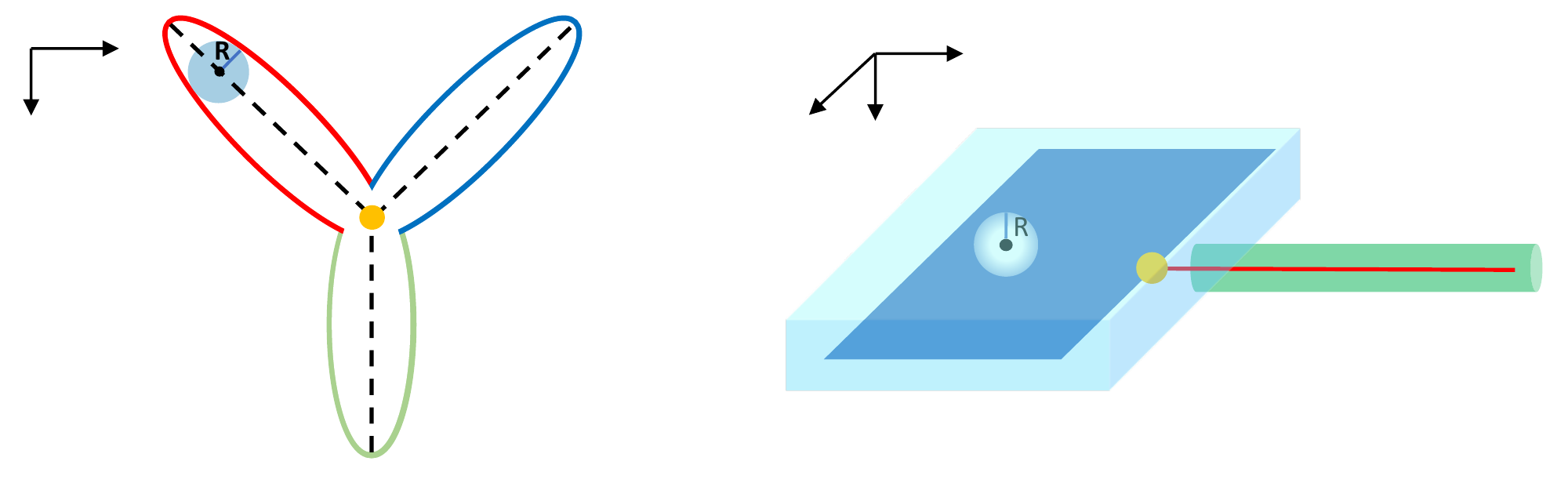}
        \put(55, -5) {\small (a)}
        \put(190, -5) {\small (b)}
    \end{overpic}
    \caption{Illustration of the topology junctions: (a) non-manifoldness and (b) change of dimensionality.}
    \label{fig:top_junctions}
\end{figure}

\para{(A) MAT topology}
This type of junction characterizes the structural (topological) clues for identifying the components of a 3D shape. See Fig.~\ref{fig:top_junctions} for an illustration and Sec. \ref{sec:struc_dec} for the detailed definitions of the junctions.
\begin{itemize}
	\item \textbf{(AA) Non-manifoldness} (Fig.~\ref{fig:top_junctions}(a)):
		Structural branchings (e.g., two legs connected to one torso) indicate a
		component-wise assembly of the object. Here, the MAT has, e.g., three
		edges meeting in a common vertex or three sheets meeting along a
		common edge.
	\item \textbf{(AB) Change of dimensionality} (Fig.~\ref{fig:top_junctions}(b)):
		Transitions from plate-like regions to tube-like regions also indicate
		part boundaries. Here, the MAT has a 2D sheet connected to a 1D edge.
\end{itemize}

\begin{figure}[!htb]
\centering
\vspace{-4mm}
    \begin{overpic}[width=\linewidth]{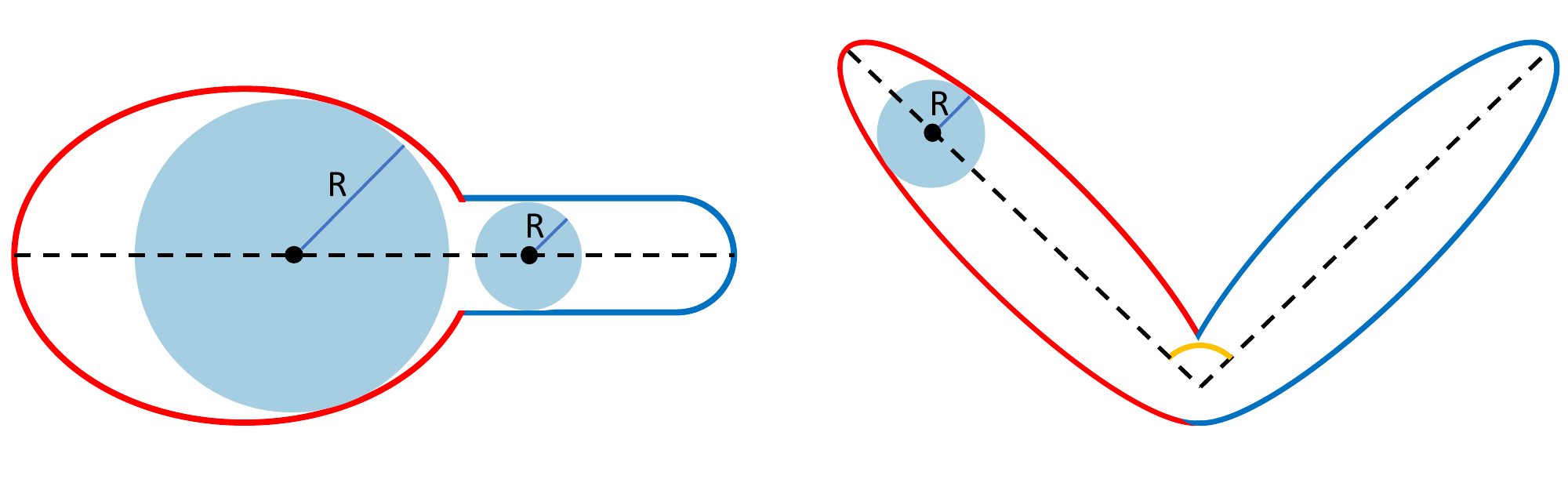}
        \put(55, -5) {\small (a)}
        \put(190, -5) {\small (b)}
    \end{overpic}
    \vspace{-5mm}
    \caption{Illustration of the geometry junctions: (a) thickness variation and (b) sharp bending.}
    \label{fig:geo_junctions}
\end{figure}

\para{(B) MAT geometry} This type of junction characterizes the geometrical clues for detecting the part boundaries of a 3D shape on the MAT. See Fig.~\ref{fig:geo_junctions} for an illustration.
\begin{itemize}
	\item \textbf{(BA) Thickness variation} (Fig.~\ref{fig:geo_junctions}(a)):
    Part boundaries are also perceived in the vicinity of concave features (valleys, incisions). On the MAT this coincides with significant variations of the sphere radii (thickness variation).
		
	\item \textbf{(BB) Sharp bending} (Fig.~\ref{fig:geo_junctions}(b)):
	Part boundaries in articulated objects are detected by observing changes	of the medial axis orientation, leading to small angles between adjacent MAT edges or sheets.
\end{itemize}

According to these insights on how two or more parts are joined together via junctions, we will give the detailed algorithm of SEG-MAT for detecting these junctions on the MAT in the next section.

\section{Method}

\begin{figure}[!htb]
\centering
    \begin{overpic}[width=\linewidth]{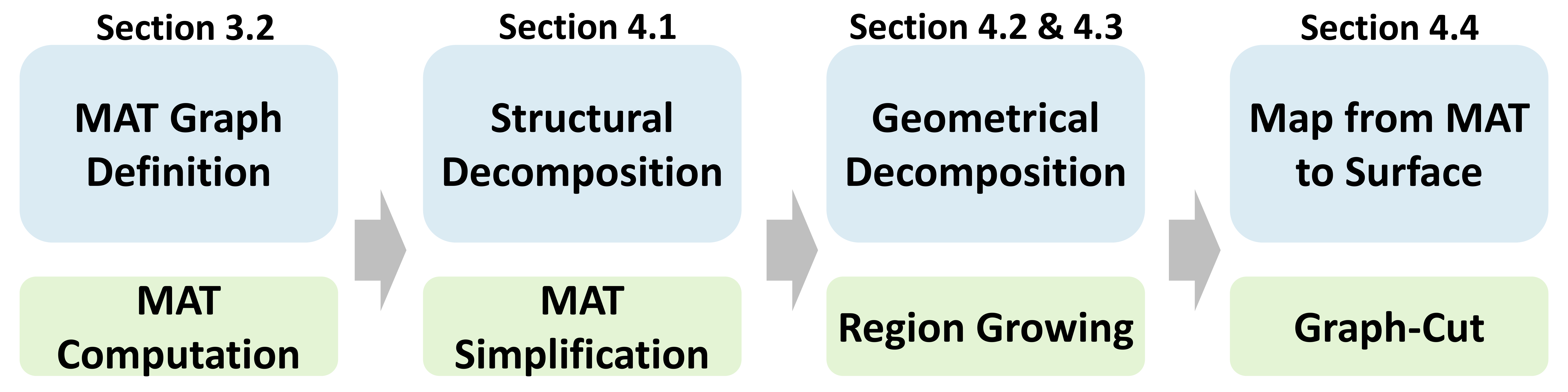}
    \end{overpic}
    \caption{An overview of the key steps of SEG-MAT. The upper row: the title of each main step; the lower row: the key technique used in each main step.}
    \label{fig:overview}
\end{figure}

We will first provide an overview of our method and then elaborate on the details of its main steps in the subsequent sections. As shown in Fig.~\ref{fig:overview}, our method has the following four main steps for segmenting a given 3D shape. (1) {\em Initialization}  (Sec.~\ref{sec:matdef}): we compute the MAT (called the {\em base MAT}) of the input 3D object and define its MAT graph.  (2) {\em Structural decomposition}  (Sec.~\ref{sec:struc_dec}): we further simplify the {\em base MAT} to obtain a {\em structured MAT} to derive the topological junctions (i.e., non-manifoldness and change of dimensionality); and cut the MAT graph into connected components along the junctions explicitly suggested by the structured MAT. This step yields a coarse segmentation result. (3) {\em Geometrical decomposition} (Sec.~\ref{sec:struc_dec} and Sec.~\ref{sec:geo_dec}): we perform region growing on each connected component by analyzing the geometrical junctions (i.e., thickness variation and sharp bending) to obtain a refined segmentation of the MAT from the coarse segmentation. (4) Finally, we map the segmentation results from the MAT domain to the surface mesh of the input shape, and further refine the segmentation results and cut boundaries (Sec.~\ref{sec:postprocess}).

\subsection{Structural decomposition}
\label{sec:struc_dec}
\begin{figure}[!htb]
\centering
    \begin{overpic}[width=0.85\linewidth]{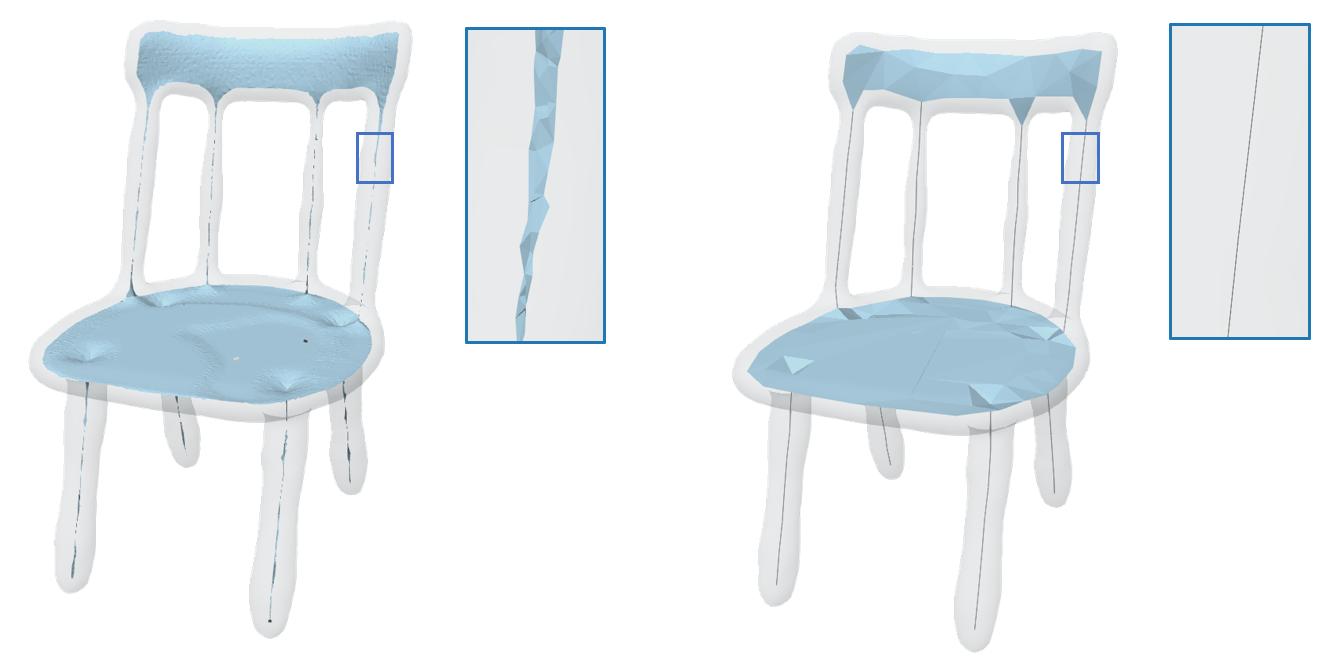}
        \put(60, 0) {\small (a)}
        \put(170, 0) {\small (b)}
    \end{overpic}
    \caption{Comparison between the base MAT and the structured MAT. (a) The medial mesh of a base MAT where the tube-like parts appear as narrow surface strips; (b) the medial mesh of the corresponding structured MAT with clean curve-sheet structures. }
    \label{fig:smat}
\end{figure}

The aim of structural decomposition is to detect topological junctions (i.e., non-manifoldness and change of dimensionality) to segment the MAT into a collection of curves and sheets, thus producing an initial coarse segmentation of the input shape. Ideally, we would like to represent a tube-like part by a skeletal curve and a plate-like part by a surface sheet, and therefore the joints between these parts can be easily identified. Nonetheless, the accurate MAT of a tube-like part may appear as a thin surface strip and the MAT of a plate-like part may contain insignificant branches due to small disturbance on the surface. We resolve this issue by using the simplification mechanism of the Q-MAT \cite{li2015q} method to further simplify the base MAT to obtain a highly simplified MAT, which is named {\em structured MAT} in this paper. A comparison between the base MAT and the {\em structured MAT} is shown in Fig.~\ref{fig:smat}.

\begin{figure}[!htb]
    \centering
    \begin{overpic}[width=\linewidth]{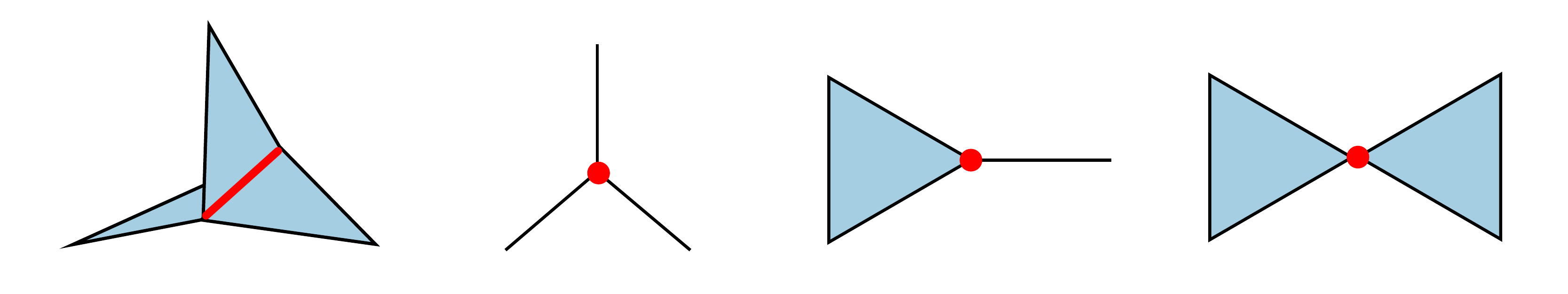}
        \put(30, -3) {\small (a)}
        \put(90, -3) {\small (b)}
        \put(145, -3) {\small (c)}
        \put(205, -3) {\small (d)}
    \end{overpic}
    \vspace{-4mm}
    \caption{Four types of joints on the structured MAT used for structural decomposition: (a) seam edge; (b) seam vertex; (c) edge-triangle vertex; (d) triangle-triangle vertex.}
    \label{fig:SMATjoints}
\end{figure}
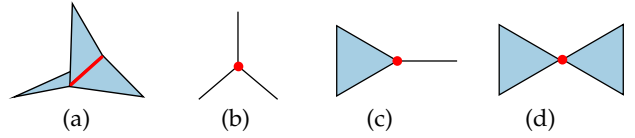

In the simplification, the triangular faces are iteratively collapsed into edges. With the error control and topology preservation mechanism of Q-MAT \cite{li2015q}, the MAT of tube-like parts will be simplified into curve skeletons, while the plate-like parts will be preserved as sheets but the insignificant branches will be removed. Please refer to the original paper \cite{li2015q} to see more detailed properties of the robustness of the simplification. In essence, the simplification produces a clean and simple curve-sheet representation, extracting an explicit structure of the base MAT. Then we identify the junctions on the {\em structured MAT} by detecting the following joints: 
\begin{itemize}

\item{Seam edge:} an edge of the {\em structured MAT} shared by at least three triangles (Fig. \ref{fig:SMATjoints} (a)).

\item{Seam vertex:}
a vertex of the {\em structured MAT} shared by at least three edges (Fig. \ref{fig:SMATjoints} (b)). 

\item{Edge-triangle vertex:}
a vertex of the {\em structured MAT} jointly shared by an edge and a triangle (Fig. \ref{fig:SMATjoints} (c)).

\item{Triangle-triangle vertex:}
a vertex of the {\em structured MAT} shared by at least two triangle faces, while the faces do not share any edge (Fig. \ref{fig:SMATjoints} (d)).
\end{itemize}

Accordingly, we decompose the structured MAT into curves and sheets using the above-mentioned joints. Then we cut the MAT graph into a set of connected components by mapping each node to the decomposed part with the closest Euclidean distance. 

\subsection{Geometrical decomposition}
\label{sec:geo_dec}
\begin{figure}[!htb]
    \begin{overpic}[width=\linewidth]{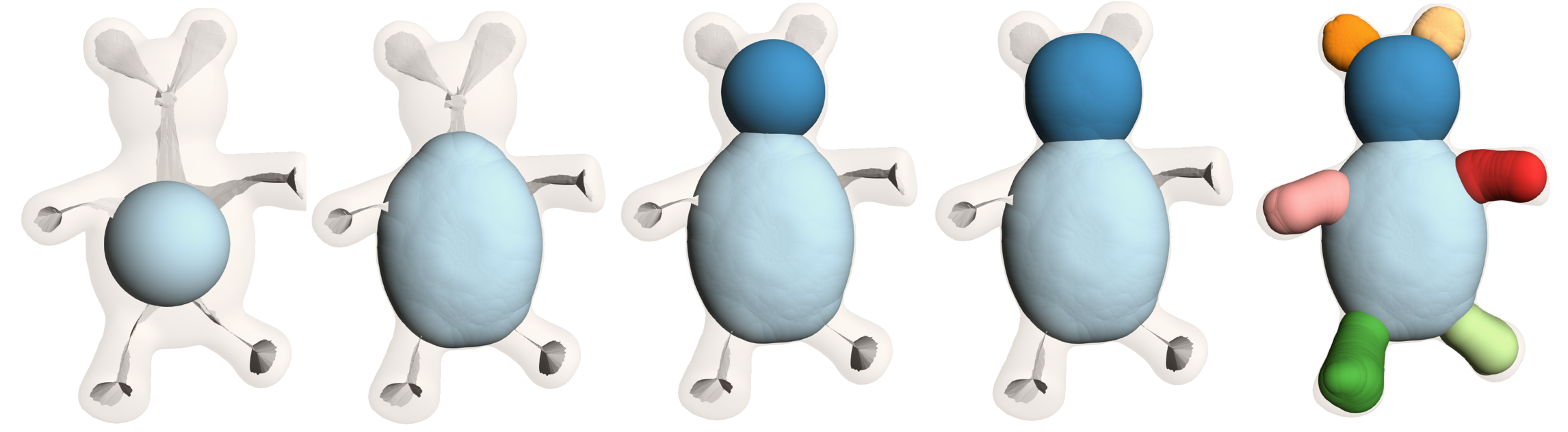}
        \put(25, -7) {\small (a)}
        \put(75, -7) {\small (b)}
        \put(125, -7) {\small (c)}
        \put(175, -7) {\small (d)}
        \put(225, -7) {\small (e)}
    \end{overpic}
    \caption{The region growing process on the MAT of a shape. (a) The first seed node with the largest radius on the MAT is selected ; (b) the first part has grown; (c) a new seed node is selected; (d) the second part has grown; (e) all the parts have grown. To clearly focus on the growing process, the MAT is givens as a single part which is not processed by the structural segmentation.}
    \label{fig:growprocess}
\end{figure}
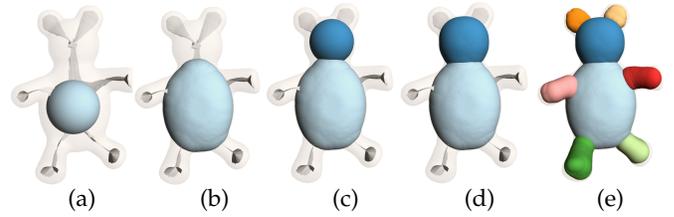

Given a coarse segmentation $\bm{\mathcal{C}}=\{ \mathcal{C}_1,  \mathcal{C}_2,...,\mathcal{C}_n\}$ by the structural decomposition, in this step, we use a region growing strategy to further segment each part $\mathcal{C}_i$ by incorporating the geometrical clues, i.e., detecting thickness variation and sharp bending junctions. The algorithm is detailed in Algorithm~\ref{alg:growingflow} with pseudo-code and the growing process is visualized in Fig.~\ref{fig:growprocess}.

\para{Region growing} The seeded region growing \cite{adams1994seeded} is a simple unsupervised method first proposed for image segmentation. It examines neighboring nodes of initial seed points and determines whether the neighbors should be added to the region. The process is iterated until there are no valid nodes that can be added. Two major problems need to be addressed when region growing is used for segmentation: how to select the seed nodes and what criteria should be adopted to characterize the growing cost. Two values need to be determined when region growing is used: the growing threshold and the minimal region threshold.

\para{Seed points selection} Determining all the seed points at the beginning is a hard problem and sometimes it needs complex computations to find the salient points or the input from users. Instead, in each iteration, we directly choose the unvisited node that has the largest medial sphere radius among all the unvisited nodes as the seed point. This simple heuristic takes advantage of the saliency of the largest radius, making the algorithm directly handle the next-thickest-part after each iteration, which is more efficient and demonstrated to be effective.

\para{Growing cost}  We propose a growing cost function on the MAT graph from two aspects. The first is the medial axis, since the radius variations and the bending angles between the triangles are important indicators of geometrical variation of a shape on the medial mesh. The second is the medial primitives (i.e., medial cones and medial slabs); they characterize the change of geometry on the level of the shape surface. Consequently, we formulate the growing cost using two terms: the {\em medial axis term} $C_{ma}$ and the {\em medial primitive term} $C_{mp}$. 

For any two adjacent nodes $N_i$ and $N_j$ of the MAT graph, the medial axis term $C_{ma}(N_i, N_j)$ is defined as
\begin{equation}
C_{ma}(N_i, N_j) =
\frac{|\overline{R}(N_i)-\overline{R}(N_j)|}{\min(\overline{R}(N_i), \overline{R}(N_j))}
+ \alpha \frac{\pi-\theta_{N_i,N_j}}{\pi}, 
\label{eq:ma_cost}
\end{equation}
where $\overline{R}(N_i)$ and $\overline{R}(N_j)$ are the average radii of $N_i$ and $N_j$; $\theta_{N_i,N_j} (\theta_{N_i,N_j}<\pi)$ is the dihedral angle if $N_i$ and $N_j$ are both face nodes, or the angle between two edges if $N_i$ and $N_j$ are both edge nodes, else 0 if $N_i$ and $N_j$ are of different types. The two terms of $C_{ma}(N_i, N_j)$ in Eq. \ref{eq:ma_cost} account for the thickness variation and bending angle between two nodes respectively. They are combined with a weight $\alpha = 0.05$, which is an empirical value based on extensive tests.

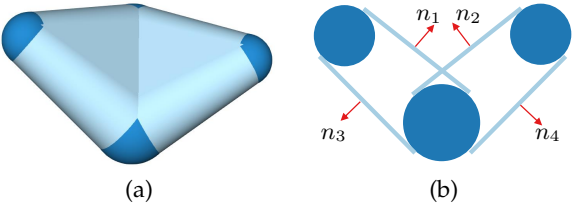
\begin{figure}[!htb]
    \centering
    \begin{overpic}[width=0.9\linewidth]{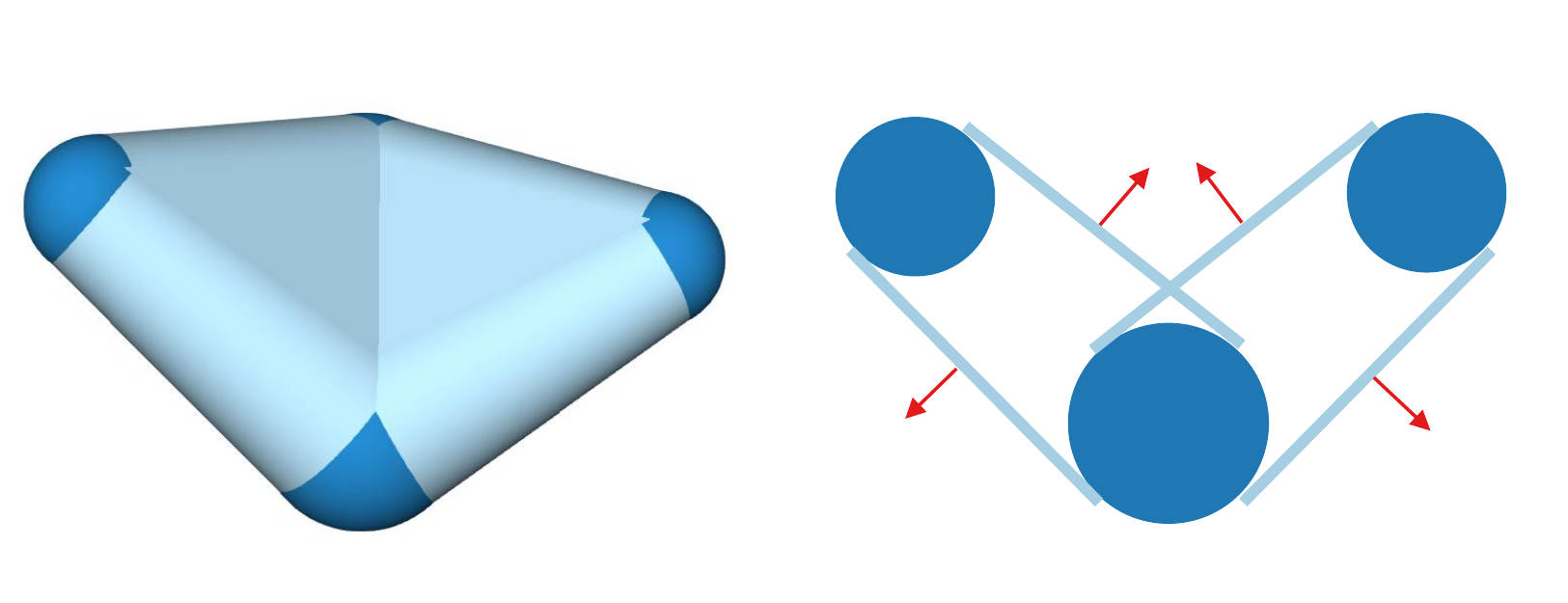} 
        \put(50, -2) {\small (a)}
        \put(165, -2) {\small (b)}
        \put(160,65){\small $n_1$}
        \put(175,65){\small $n_2$}

        \put(124,20){\small $n_3$}
        \put(205,20){\small $n_4$}
    \end{overpic}
    \caption{Two-side primitive angles. (a) Two joining slabs; (b) illustration of two angles (given by two pairs of normals $(n_1,n_2)$ and $(n_3,n_4)$ ) formed on the two sides. }
    \label{fig:slab_angle}
\end{figure}

To define the primitive term $C_{mp}$, we first recall the medial primitives in an MAT (see Sec.~\ref{sec:matdef}), namely, the {\em medial cone} defined by an edge and the {\em medial slab} defined by a triangle face of the medial mesh. As shown in Fig.~\ref{fig:slab_angle}, the union of two adjacent primitives will form two angles on the two sides. These angles are concentrated expressions of the thickness change and sharp bending on the surface; that is, either thickness change or sharp bending will lead to these angles between primitives. The medial primitive term $C_{mp}$ is defined as
\begin{equation}
C_{mp}(N_i, N_j)= \frac{\angle(N_i^+,N_j^+)+\angle(N_i^-,N_j^-)}{2\pi},
\end{equation}
where $\angle(N_i^+,N_j^+)$ and $\angle(N_i^-,N_j^-)$ denote the primitive angles computed by two pairs of normals $(n_1,n_2)$ and $(n_3,n_4)$ (see Fig.~\ref{fig:slab_angle}) of the two sides respectively. Finally, the mesh term $C_{ma}$ and the primitive term $C_{mp}$ are combined as follows to derive the final growing cost function:
\begin{equation}
C(N_i, N_j)=min\{C_{ma}(N_i, N_j), \lambda C_{mp}(N_i, N_j)\}.
\label{eq:growingFull}
\end{equation}
Here $\lambda$ is the weight factor fixed at $1.5$ empirically. Since both $C_{ma}$ and $C_{mp}$ are the expressions of geometrical change from two perspectives, using the {\em min} operator results in higher reliability; that is, only if the values of the two (weighed) terms both exceed a threshold, a segmentation is suggested.

\para{Growing threshold} The growing threshold $\delta$ is an important parameter to gauge the growing process. The node $N_j$ will be added into the region of $N_j$ only if $C(N_i, N_j)\le\delta$. 
\begin{wrapfigure}{r}{0.4\linewidth}
\vspace{-2mm}
\hspace{-2mm}
\includegraphics[width=\linewidth]{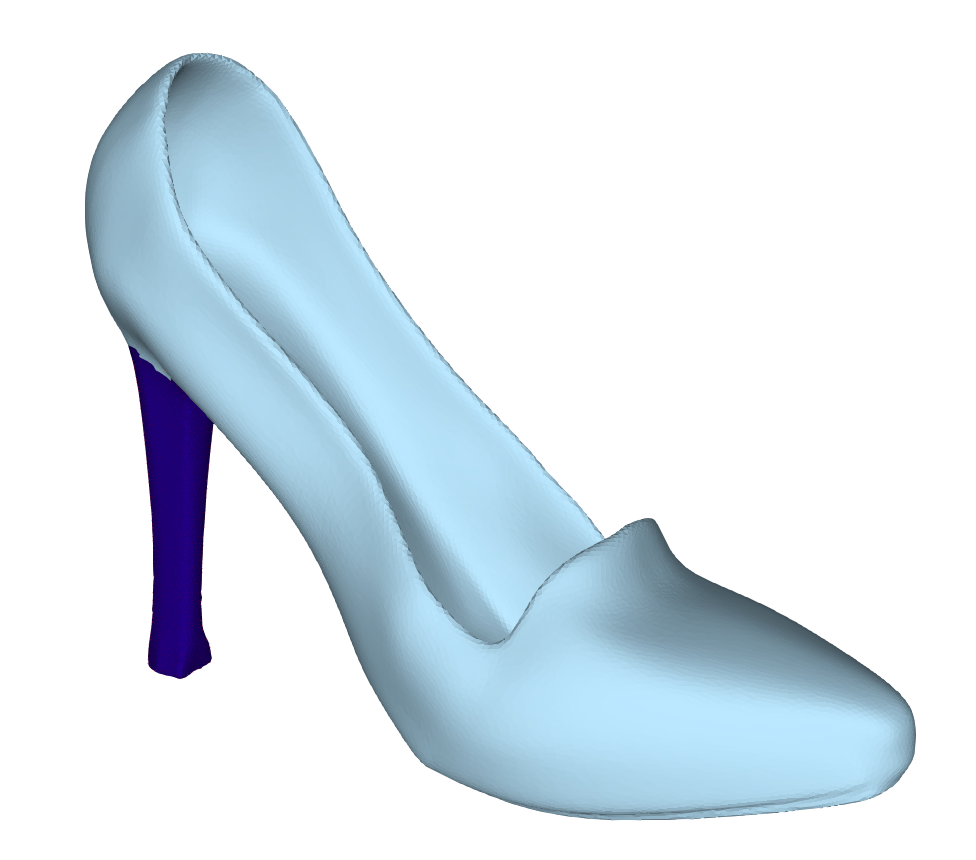}
\vspace{-2mm}
\end{wrapfigure}  
It is worth noting that, people's sensitivity to geometrical variation for different forms of parts differs. A key observation is, for some ``thin'' parts, although the geometry changes drastically on this part, people still usually deem it as a single part. For example, the upper piece of the shoe model in the inset, despite being highly non-convex, should be deemed as a whole part. Based on this insight, given a set of curves and sheets produced by structural decomposition, we first measure how ``thin'' a component $\mathcal{C}_i$ is by 
\begin{equation}
    \rho(\mathcal{C}_i)=\frac{S(\mathcal{C}_i)}{R_{max}(\mathcal{C}_i)},
    \label{eq:tubeplatecheck}
\end{equation}
where $S(\mathcal{C}_i)$ is the length of $\mathcal{C}_i$ if it is a curve segment of {\em structured MAT} or the square root of the area of $\mathcal{C}_i$ if it is a sheet,  and $R_{max}(\mathcal{C}_i)$ is the largest radius on $\mathcal{C}_i$. Intuitively, it measures how ``thin'' a component is in terms of the ratio of the shape expansion $S(\mathcal{C}_i)$ to the shape thickness expressed by radius $R_{max}(\mathcal{C}_i)$. If $\mathcal{C}_i$ is a thin tube or plate, $\rho(\mathcal{C}_i)$ should be fairly large. 

For the thin parts, we propose to use a larger threshold to encourage merging so as to maintain the integrity of these tubes or sheets by giving them higher tolerance to geometrical change. Therefore, the growing threshold is automatically adjusted by
\begin{equation}
    \delta=\delta_0 \cdot \sigma(\log(\rho({\mathcal{C}_i})).
\end{equation}

Here $\rho({\mathcal{C}_i})$ is the thinness of $\mathcal{C}_i$, and $\sigma(x)$ is an indicator function where $\sigma(x)=x$ if $x\ge 3$ or it is $1$. This indicator function is to filter out the bulky parts that are not thin enough and thus not considered for threshold adjustment. $\delta_0$ is the original threshold given by a user; note that $\delta \ge \delta_0$. Our tests have shown that this automatic threshold adjustment scheme works effectively for the segmentation of general shapes. The default value of $\delta_0$ is set to 0.015 in this paper, but can be tuned according to users' requirements in practice, see Sec.~\ref{subsec:parameterAnalysis}.

\para{Minimal region threshold} As a necessary parameter of the region growing algorithm, minimal region threshold $\eta$ is to guarantee that no region is smaller than this threshold to avoid fragmented segmentation. Here we consider a generated part as {\em negligible} if the ratio of the number of its nodes to the number of all the nodes of the entire graph is less than $\eta$. The negligible parts will be merged into one of its adjacent parts. We set $\eta$ to $0.2\%$ and evaluate it in Sec.~\ref{subsec:parameterAnalysis}.

\begin{algorithm}[!htb]
\KwIn{ MAT Graph $G=\{\{N_i\}, \mathcal{E}\}$, which is cut into a set of components $\bm{\mathcal{C}}=\{ \mathcal{C}_1,  \mathcal{C}_2,...,\mathcal{C}_n\}$ }
$\slash\slash$ The growing cost $\delta_0$ and the least node ratio $\eta$ are pre-defined. \\
\KwOut{Fine-grained MAT components $\bm{\mathcal{R}}=\{ \mathcal{R}_1,  \mathcal{R}_2,...,\mathcal{R}_m\}$ }
Build an empty Queue $Q$;\\
\Repeat{all nodes are visited}
    {
        Initialize a new part $\mathcal{R}'$;\\
        $seed$ = the unvisited node $N$ with the largest $\overline{R}(N)$;\\
        $Q.push(seed)$;\\
        Mark $seed$ as visited;\\
        \While{Q is not empty}
        {
           $N_i\gets Q.pop()$;\\
           \text{Add $N_i$ to part $\mathcal{R}'$};\\
           \For{each $N_j$ where ($N_i$,$N_j$) $\in \mathcal{E}$}
        {
           \If{$N_j$ is unvisitied  \textbf{and} $C(N_i,N_j)<\delta_0 \cdot\sigma(\log(\rho(\mathcal{C}_i))$}
           {
              $Q.push(N_j)$;\\
              Mark $N_j$ as visited;\\
           }
        }
        }
        \eIf{the node number ratio of $\mathcal{R}' >= \eta$}
        {
           \For{each unvisited node $N_k$}
           {
           \If{$N_k$ should be swallowed by $\mathcal{R}'$}
           {
           Add $N_k$ to the part $\mathcal{R}'$;\\
           Mark $N_k$ as visited;\\
           }
           }
           Add $\mathcal{R}'$ to $\bm{\mathcal{R}}$;\\
        }
        {
           Mark the nodes in $\mathcal{R}'$ as negligible nodes; 
        }
    }

\caption{Region Growing on the MAT.}
\label{alg:growingflow}
\end{algorithm}

\subsection{Swallowing}
\label{sec:swallowing}
As aforementioned, the MAT is sensitive to the boundary noise; small perturbations to the shape boundary will result in numerous long and unstable spikes on the MAT. For the segmentation task, these spikes should not be treated as valid parts. We therefore propose an efficient method, called {\em swallowing}, to handle these spikes in the process of region growing.

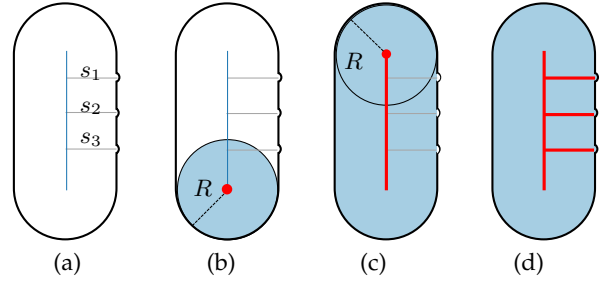
\begin{figure}[!htb]
    \vspace{-3mm}
    \centering
    \begin{overpic}[width=\linewidth]{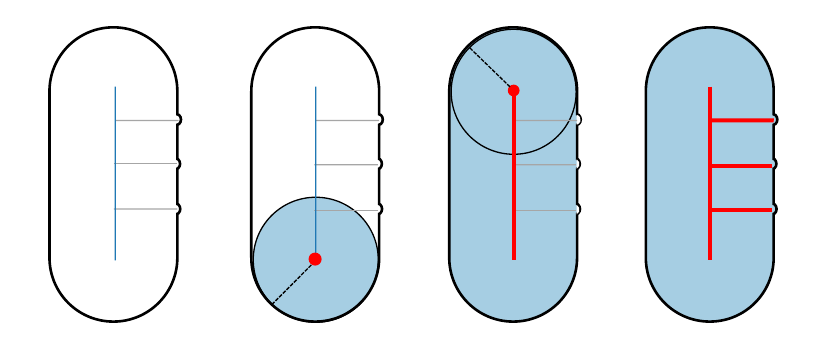}
        \put(40,70){\small $s_1$}
        \put(40,57){\small $s_2$}
        \put(40,45){\small $s_3$}
        \put(83,25){\small $R$}
        \put(140,73){\small $R$}
        \put(30, -3) {\small (a)}
        \put(88, -3) {\small (b)}
        \put(146, -3) {\small (c)}
        \put(204, -3) {\small (d)}
    \end{overpic}
    \vspace{-5mm}
    \caption{The swallowing process. (a) The MAT of a shape with spikes ($s_1$, $s_2$ and $s_3$ colored in gray) due to boundary noise; (b) start growing from a seed node (red); (c) the grown part (colored in blue) with its corresponding growing trace (colored in red), and the region includes $s_1$, $s_2$ and $s_3$; (d) the spikes are swallowed by the grown region.} 
    \label{fig:swallow}
\end{figure}

Since the growing starts from the node with the largest radius among the unvisited nodes, a newly generated volumetric region can be viewed as the trunk of this part. As shown in Fig. \ref{fig:swallow}, a generated region can include most of the unstable spikes inside this trunk part. Therefore, once a volumetric region is generated, we find out the unvisited nodes as well as the negligible nodes whose corresponding faces or edges in the MAT are enclosed in or intersect with any spheres in the generated region. Then we merge these nodes into the generated region and mark them as visited on the graph. This strategy makes the growing process only focus on the backbone of the MAT and swallow the branching spikes, making our algorithm insensitive to shape noise (see the validation in Sec. \ref{sec:anti-noise}) and robust to the instability of the MAT. 

\subsection{Transferring segments to surface}
\label{sec:postprocess}
Before transferring the segmentation results from the MAT to the surface, we adopt a post-processing step to further refine the results. Following Kaick et al.~\cite{wcseg}, we compute a histogram of the radius distribution of each part, and measure the Earth Mover's Distance (EMD) \cite{rubner2000emd} between two radius distributions; two adjacent parts will be merged iteratively if they have matched radius distribution. This step employs more global information beyond the local region growing to further avoid over-segmentation. 

In order to obtain smooth cut boundaries, we do not directly map the segmentation results from the MAT graph to the surface mesh. Consider the dual graph of the surface mesh $G=\{\mathcal{F}, \mathcal{N}\}$, where $\mathcal{F}$ denotes the set of nodes representing the triangle faces and $\mathcal{N}$ the set of edges between neighboring faces. We formulate a $k$-way graph-cut optimization problem to find the optimal MAT segment for each triangle face by minimizing the following energy:

\begin{equation}
    E(L)=\sum\limits_{f\in\mathcal{F}} E_d(l_f) +\omega \sum\limits_{(f,g)\in\mathcal{N}} E_s(l_f, l_g),
\label{eq:graphcut}
\end{equation}
where $E_d$ is the data term to penalize the assignment of a face node $f$ to the label $l_f$, which is measured by the closest distance between $f$ and a corresponding MAT segment. Specifically, it is computed as the Euclidean distance between the centroid of a face and its closest medial sphere of an MAT segment, and then normalized by the length of the diagonal of the object's bounding box. The second term $E_s(l_f, l_g)$ is the smoothness term defined as 
\begin{equation}
    E_s(l_f, l_g)=
    \begin{cases}
        0 &l_f=l_g\\
        \min(\frac{\phi(f,g)}{\pi},1) &l_f \ne l_g,
    \end{cases}
\end{equation}
where $\phi(f,g)$ is the exterior dihedral angle between face $f$ and face $g$. The parameter $\omega=0.3$ balances the data fitting term and the smoothness term. We use the method of \cite{delong2012fast} to efficiently solve this combinatorial optimization problem. This step optimizes the correspondences between the MAT segments and the surface triangles, encouraging the cut boundaries to smoothly pass through the concave valleys of the surface mesh.

\section{Experimental Results}
In this section, we conduct extensive comparisons and discussions to evaluate our algorithm. We use the Princeton Segmentation Benchmark (PSB) \cite{chen2009benchmark} to compare with rule-based methods, and use a subset of the testing data of ShapePFCN \cite{kalogerakis20173d}, which is collected from ShapeNet \cite{chang2015shapenet}, to compare with learning-based methods. The ShapeNet models are remeshed into watertight manifold surfaces \cite{huang2018manifold} first for robust MAT computation.

Following the error control metric of Q-MAT \cite{li2015q}, the base MAT is computed by setting the average slab quadratic error (SQE) to $0.5\%$ and for the structured MAT it is set to $3\%$, both relative to the length of the diagonal of the bounding box. All the experiments are performed on a machine with Intel Core i7-7700K 4.2GHz CPU and 32GB RAM.

\begin{figure}[!htb]
    \begin{overpic}[width=\linewidth]{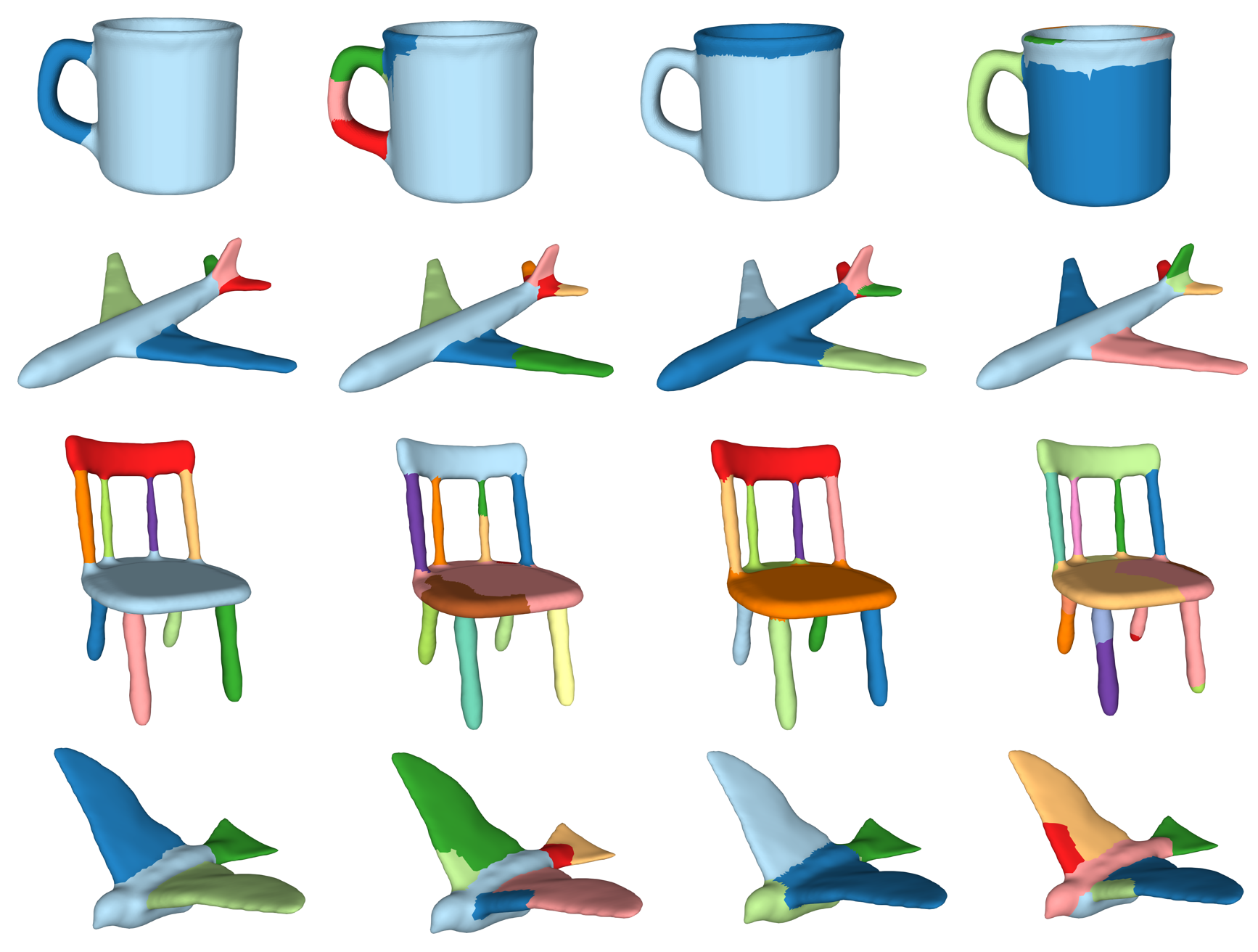}
        \put(15, -7) {\small SEG-MAT}
        \put(85, -7) {\small WCSeg}
        \put(150, -7) {\small RandCuts}
        \put(220, -7) {\small SDF}
    \end{overpic}
    \vspace{-2mm}
    \caption{Qualitative comparison with representative rule-based geometrical segmentation methods. }
    \label{fig:visual_cmp_geo}
\end{figure}

\begin{figure*}[!htb]
    \begin{overpic}[width=\linewidth]{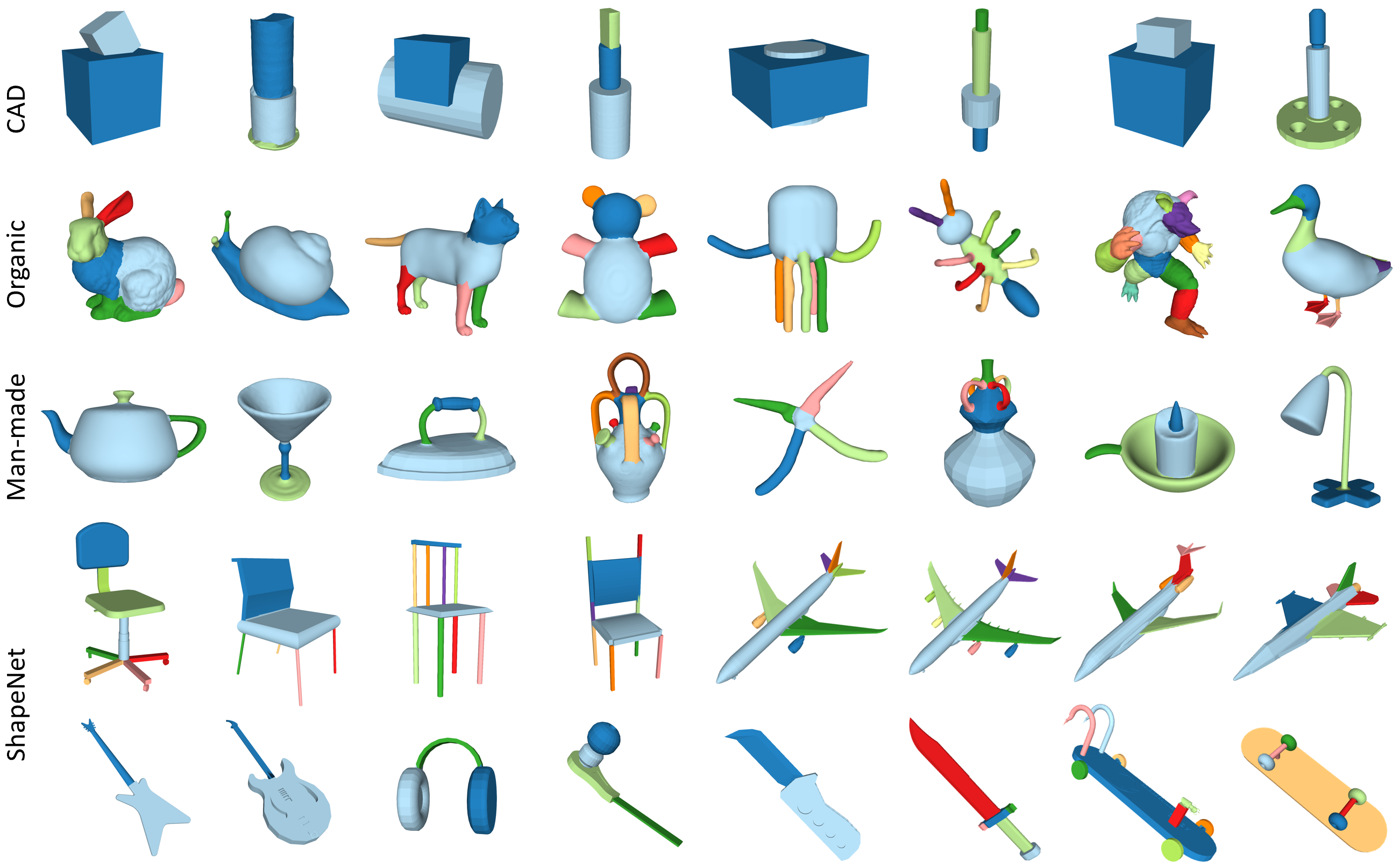}
    \end{overpic}
    \vspace{-4mm}
    \caption{Segmentation results of SEG-MAT for various shapes. We always use consistent parameter settings for different shapes and datasets; the number of parts of each shape is determined by the algorithm automatically.}
    \label{fig:overview_results}
\end{figure*} 

We show our segmentation results for a variety of 3D shapes in Fig.~\ref{fig:overview_results}. We use the same parameter settings for shapes of different categories and different datasets. Our method is able to segment various types of shapes and determine the number of parts automatically using consistent settings, which demonstrates the flexibility and robustness of our method.

\begin{table}[!htb]
\caption{Comparison with rule-based geometrical method on PSB dataset for each object category. Entries represent the rank of the method based on the Rand Index evaluation metric (1 is the best, and 7 is the worst).}
\label{tab:percategoryRI}
\centering
\setlength{\tabcolsep}{1.5mm}
\begin{tabular}{c|ccccccc}\hline
          & SEG        & WC         & Rand       & Shape & Norm       & Core  & Rand  \\
          & MAT        & Seg        & Cuts       & Diam  & Cuts       & Extra & Walks \\\hline
Human     & 2          & \textbf{1} & 3          & 4     & 5          & 6     & 7     \\
Cup       & \textbf{1} & 2          & 3          & 7     & 4          & 5     & 6     \\
Glasses   & \textbf{1} & 4          & 2          & 5     & 3          & 6     & 7     \\
Airplane  & \textbf{1} & 2          & 4          & 3     & 5          & 6     & 7     \\
Ant       & \textbf{1} & 2          & 4          & 3     & 5          & 6     & 7     \\
Chair     & \textbf{1} & 3          & 7          & 4     & 2          & 6     & 5     \\
Octopus   & 2          & \textbf{1} & 6          & 3     & 5          & 4     & 7     \\
Table     & \textbf{1} & 2          & 7          & 5     & 3          & 6     & 4     \\
Teddy     & 2          & 3          & \textbf{1} & 4     & 6          & 5     & 7     \\
Hand      & 2          & 3          & \textbf{1} & 6     & 5          & 4     & 7     \\
Plier     & \textbf{1} & 2          & 4          & 7     & 5          & 3     & 6     \\
Fish      & 2          & \textbf{1} & 5          & 3     & 6          & 4     & 7     \\
Bird      & \textbf{1} & 2          & 3          & 4     & 6          & 5     & 7     \\
Armadillo & 3          & 2          & \textbf{1} & 4     & 6          & 7     & 5     \\
Bust      & 3          & 2          & \textbf{1} & 4     & 6          & 5     & 7     \\
Mech      & 3          & 2          & 6          & 4     & \textbf{1} & 7     & 5     \\
Bearing   & \textbf{1} & 3          & 4          & 2     & 5          & 7     & 6     \\
Vase      & 3          & 2          & \textbf{1} & 5     & 6          & 4     & 7     \\
FourLeg   & \textbf{1} & 2          & 4          & 3     & 5          & 6     & 7     \\\hline
Overall   & \textbf{1} & 2          & 3          & 4     & 5          & 6     & 7     \\\hline
\end{tabular}
\end{table}

\begin{table}[!htb]
\caption{Analysis of the trade-off among computation time, segmentation quality and usability. (*: whether the numbers are obtained by tuning the parameters for each shape or category separately; \#: the computation time is measured when the MAT is given, while the MAT computation time for one shape using Q-MAT \cite{li2015q} is about 4s.) }
\label{tab:trade-off}
\centering
\setlength{\tabcolsep}{1.5mm}
\begin{tabular}{c|cccc}\hline
                     & SEG-MAT        & Shu et al. & WCSeg      & Randcut \\\hline
Avg. Computation Time  & \textbf{4.7s$ ^\#$}  & 216.3s  & 127.4s    & 78.8s    \\
Rand Index           & \textbf{0.114} & 0.118      & 0.123       & 0.157   \\
Per-category Tuning$^*$    & \textbf{No}    & Yes        & \textbf{No} & Yes    \\\hline
\end{tabular}
\vspace{-3mm}
\end{table}

\subsection{Comparison with rule-based methods}
\label{sec:cmp-rule}
We first qualitatively and quantitatively evaluate our method with comparisons to rule-based segmentation methods that use geometrical analysis for single shape segmentation on the PSB \cite{chen2009benchmark}. These methods include weakly convex segmentation (WCSeg) \cite{wcseg}, randomized cuts (RandCuts) \cite{randcut}, shape diameter function (SDF) \cite{sdf2008}, normalized cuts (NormCuts) \cite{randcut}, core extraction (CoreExtra) \cite{coreextra}, random walks (RandWalks) \cite{randomwalk}, primitive fitting (FitPrim) \cite{fitprim}, and K-Means \cite{kmeans}. Fig. \ref{fig:visual_cmp_geo} shows the qualitative comparison results with some representative methods. It can be observed SEG-MAT gives better segmentation results that are more consistent with human's perception.

\begin{table*}[!htb]
\caption{Comparison with rule-based segmentation algorithms using four metrics defined on the PSB dataset: Rand Index, Cut Discrepancy, Hamming Distance and Consistency Error. RI: Rand Index; CD: Cut Discrepancy; GCE: Global Consistency Error; LCE: Local Consistency Error; HD: Hamming Distance; HD-Rm: Hamming missing rate; HD-Rf: Hamming false alarm rate. The metrics are scaled by 1000 for better readability.}
\label{tab:psb_overall_eval}
\centering
\setlength{\tabcolsep}{5.0mm}
\begin{tabular}{c|ccccccc}\hline
          & RI             & CD             & GCE           & LCE           & HD        & HD-Rm     & HD-Rf     \\\hline
SEG-MAT    & \textbf{114.6} & \textbf{212.1} & \textbf{95.9} & 63.3          & \textbf{117.1} & \textbf{115.1} & 119.1          \\
WCSeg     & 122.7          & 214.6          & 98.1          & \textbf{62.4} & 118.4          & 120.6          & \textbf{116.3} \\
RandCuts  & 157.4          & 274.1          & 122.1         & 71.1          & 136.2          & 142.9          & 129.5          \\
ShapeDiam & 175.7          & 274.7          & 129.6         & 82.3          & 166.5          & 186.7          & 146.2          \\
NormCuts  & 178.0          & 299.4          & 150.4         & 97.0          & 175.6          & 183.1          & 168.1          \\
CoreExtra & 210.8          & 375.2          & 135.0         & 86.3          & 169.4          & 126.2          & 212.5          \\
RandWalks & 229.7          & 384.5          & 175.1         & 103.3         & 208.2          & 205.6          & 210.9          \\
FitPrim   & 216.4          & 349.4          & 217.2         & 143.0         & 241.1          & 285.0          & 197.1          \\
KMeans    & 254.9          & 420.0          & 249.3         & 166.8         & 277.1          & 334.2          & 220.1         \\\hline
\end{tabular}
\end{table*}

As shown in Table \ref{tab:psb_overall_eval}, we use four metrics defined by the PSB \cite{chen2009benchmark}, i.e.,  Rand Index, Cut Discrepancy, Hamming Distance and Consistency Error, for quantitative comparison. It can be seen that our algorithm is comparable or better than the state-of-the-art methods, but is about one order of magnitude faster (see the next paragraph). Following the metrics in \cite{chen2009benchmark} and \cite{wcseg}, we show in Table \ref{tab:percategoryRI} the rankings of these algorithms for each shape category in terms of the Rand Index score. SEG-MAT performs the best in 10 categories out of 19 categories, while none of the other methods win in more than 5 categories.

We also jointly examine the average running time, performance and usability for different methods on the PSB dataset, of which results are shown in Table~\ref{tab:trade-off}. The usability is embodied by whether the results are obtained by using category-specific or shape-specific parameters. Here we also include an unsupervised learning-based method \cite{shu2016unsupervised} that aggregates multiple shape signatures for comparison. We can see the other methods that need to extract low-level geometrical descriptors usually take a longer time to process a single 3D shape. However, SEG-MAT uses significantly less computation time and consistent parameters for different shapes and categories but achieves the best performance.

\begin{figure*}[!htb]
\centering
    \begin{overpic}[width=\linewidth]{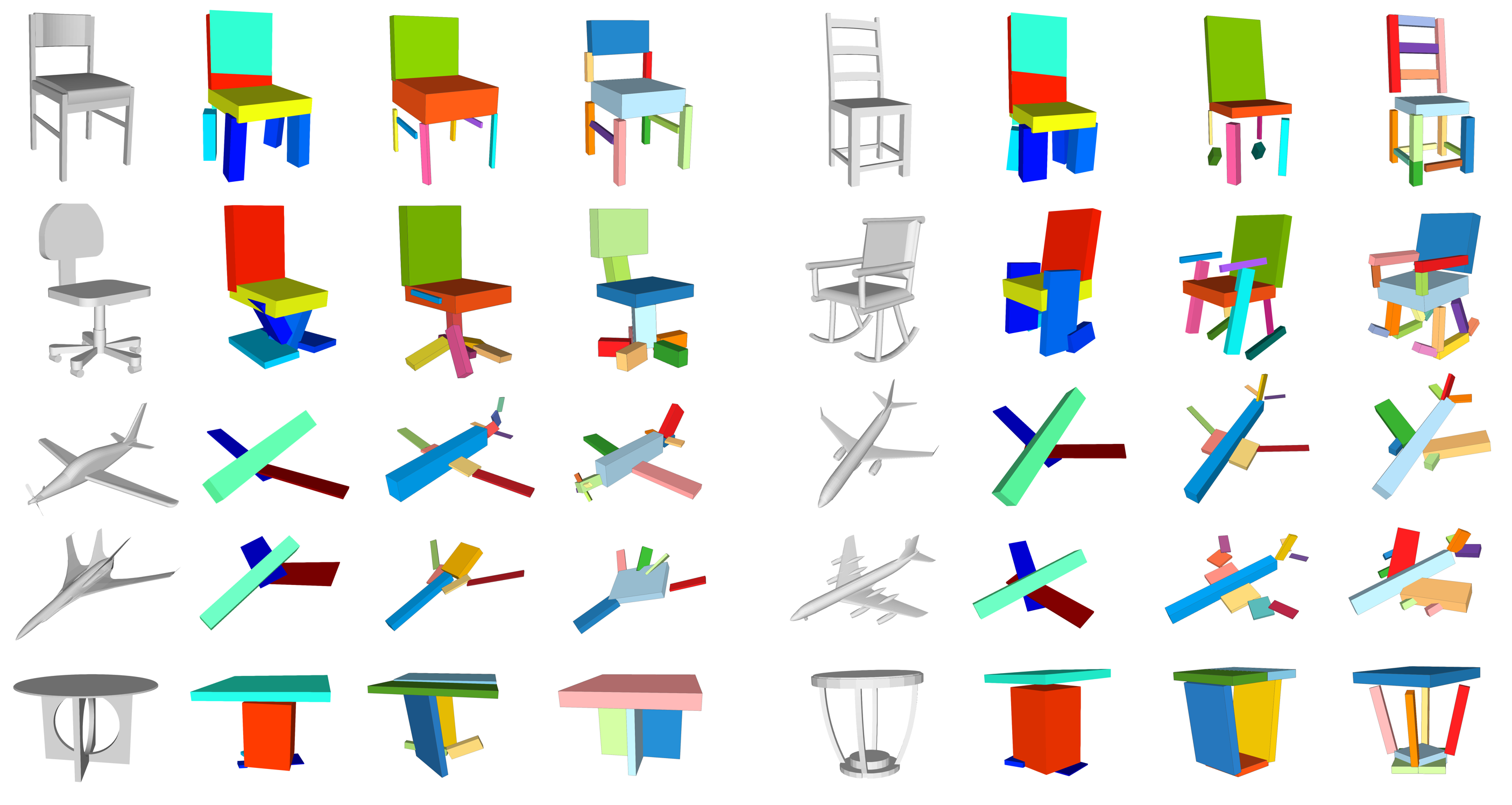}
        \put(20, -7) {\small Input}
        \put(70, -7) {\small Tulsiani et al.}
        \put(140, -7) {\small Sun et al.}
        \put(200, -7) {\small SEG-MAT}
        \put(290, -7) {\small Input}
        \put(335, -7) {\small Tulsiani et al.}
        \put(410, -7) {\small Sun et al.}
        \put(460, -7) {\small SEG-MAT}
    \end{overpic}
    \vspace{0mm}
    \caption{Qualitative comparison with unsupervised learning methods for primitive-based approximation and part-based segmentation. }
    \label{fig:prim_results}
\end{figure*} 

\begin{table}[!htb]
\caption{Quantitative comparison with the unsupervised learning-based methods using ShapeNet \cite{chang2015shapenet} data. We report the IoU and the Chamfer Distance between the generated primitive-based representations and the original shapes.}
\label{tab:prim_cmp}
\setlength{\tabcolsep}{1.5mm}
\begin{tabular}{c|ccc|ccc}\hline
                & \multicolumn{3}{c|}{IoU} & \multicolumn{3}{c}{Chamfer Distance}\\\hline
                & Airplane       & Chair          & Table          & Airplane        & Chair           & Table           \\\hline
Tulsiani et al. & 0.271          & 0.313          & 0.344          & 0.0531          & 0.0506          & 0.0537          \\
Sun et al.      & 0.445          & 0.471          & 0.409          & 0.0301          & \textbf{0.0338} & 0.0434          \\
SEG-MAT         & \textbf{0.486} & \textbf{0.474} & \textbf{0.436} & \textbf{0.0260} & 0.0395          & \textbf{0.0347}\\\hline
\end{tabular}
\end{table}

\begin{table}[!htb]
\caption{Elaboration on the key differences between semantic segmentation and geometrical segmentation methods.}
\begin{tabular}{p{2cm}|p{2.7cm}|p{2.7cm}}
\hline
                       & Semantic learning                              & Geometrical analysis
               \\\hline
Application condition  & A group of similar shapes in the same category & Arbitrary single shape                                              \\\hline
Produce language label & Yes                                            & No                                                        \\\hline
Need human annotation  & Yes                                            & No                                                        \\\hline
Segmentation criteria  & Consistency with pre-defined semantic labels    & Minima rule (parts are weakly convex)                     \\\hline
Number of segments     & Same within a category                         & Different within a category                               \\\hline
Ground truth           & Only one                                       & One or more                                               \\\hline
Evaluation metrics     & Labeling accuracy                              & Convexity, part-based abstraction error, Rand Index, etc.\\\hline
Representative methods & ShapePFCN\cite{kalogerakis20173d}, Xu et al. \cite{xu2017directionally}   & Rule-based: WCSeg\cite{wcseg}, RandCut\cite{randcut}  Learning-based: Su et al. \cite{abstractionTulsiani17}, Sun et al. \cite{sun2019hierarchy} \\\hline 
\end{tabular}
\label{tab:geo-seman-diff}
\end{table}

\subsection{Comparison with learning-based methods}
\label{sec:cmp-learning}
\para{Comparison with unsupervised learning methods} The unsupervised learning methods do not require manually annotated labels. The segmentation also relies on the intrinsic geometrical properties of a shape rather than pre-defined labels and the results follow the minima rule \cite{hoffman1984parts}. Therefore, learning-based methods of this category have the same goal as our algorithm and can be included for comparison. Recently, Tulsiani \emph{et al.} \cite{abstractionTulsiani17} and Sun \emph{et al.} \cite{sun2019hierarchy} propose unsupervised methods for structural analysis of 3D shapes. They learn to parse a 3D shape into a set of cuboid primitives and induce a segmentation by the projection of the predicted primitives onto the original shape. We follow their metrics for evaluation: we use a tight cuboid primitive (i.e., minimal oriented bounding box, see Sec.~\ref{sec:more_applications} for the detail) to represent each part; then a shape is rebuilt by a set of primitives and we evaluate the reconstruction error. 

Fig.~\ref{fig:prim_results} shows qualitative comparison results with these methods, where our results are more structurally meaningful and able to capture more detailed components. Table~\ref{tab:prim_cmp} reports the intersection-over-union (IoU) and the Chamfer Distance between the primitive-based representations and the target shapes. We obtain smaller reconstruction errors in most categories, which means our method can achieve better approximation accuracy as well as segmentation quality. Note that their methods need to be trained separately on each category using a large amount of data, while we use consistent parameter settings for all categories.

\para{Discussion of supervised learning (semantic) methods} \label{sec:semantic_discussion} The supervised learning methods for semantic segmentation have different applications, goals and evaluation metrics compared to the methods based on geometrical analysis. At the beginning of this paper, we give a comparison between the segmentation results of semantic and geometrical methods in Fig.~\ref{fig:two-cat-diff}, which shows semantic segmentation targets integral semantic clusters but does not focus on part instances. In Table~\ref{tab:geo-seman-diff}, we further elaborate on their differences from various aspects to give more insights.

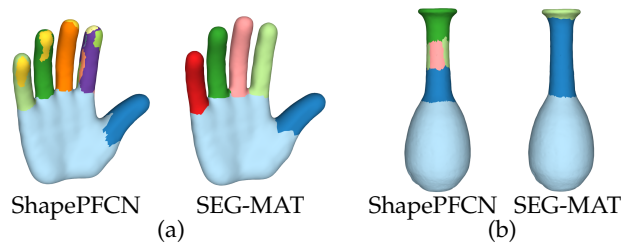
\begin{figure}[!htb]
\centering
    \begin{overpic}[width=0.9\linewidth]{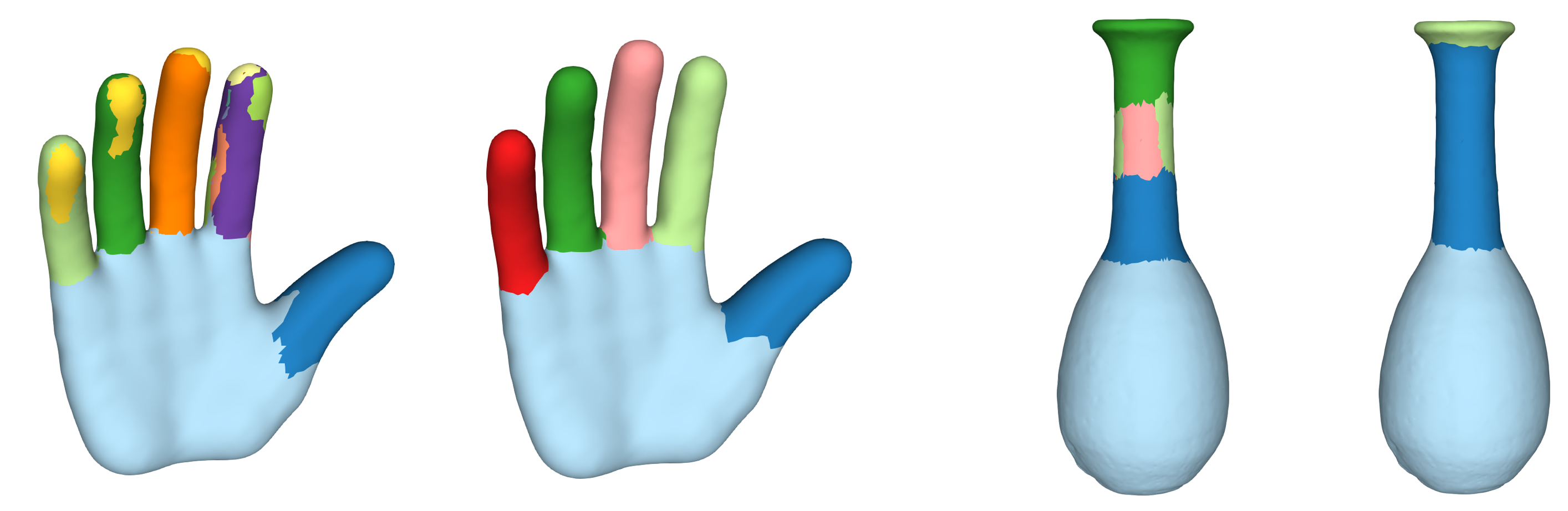}
        \put(5,-6) {\small ShapePFCN}
        \put(75, -6) {\small SEG-MAT}
        \put(60, -15) {\small (a)}
        \put(140, -6) {\small ShapePFCN}
        \put(195, -6) {\small SEG-MAT}
        \put(185, -15) {\small (b)}
    \end{overpic}
    \vspace{4mm}
    \caption{Visual comparison between ShapePFCN (supervised learning method for semantic segmentation) and SEG-MAT.}
    \label{fig:cmp_pfcn}
\end{figure}

Specifically, we discuss the ShapePFCN \cite{kalogerakis20173d}, a state-of-the-art supervised learning method. The ShapePFCN aims to label a 3D shape with pre-defined semantic meanings and needs annotated data for training. With strong supervision, ShapePFCN achieves superior segmentation results with interpretable language labels. However, some issues of the semantic segmentation methods cannot be neglected. These methods rely on the fixed semantics of the training data and focus on minimizing the labeling error, but barely consider the explicit geometrical properties of a 3D shape. This can lead to high labeling accuracy, but sometimes the results are counter-intuitive. As shown in Fig.~\ref{fig:cmp_pfcn} (a), for the hand model, ShapePFCN produces some fragmented pieces on the surface where the geometry does not change appreciably. Another interesting phenomenon is, even the vase shown in Fig.~\ref{fig:cmp_pfcn} (b) does not have a handle, ShapePFCN still labels a handle part to match the semantics of the training data. Our method works well for segmenting these two simple models, with the help of the clear geometrical differentiation.

\subsection{Comparison with skeleton-based methods}
\begin{figure}[H]
\centering
    \begin{overpic}[width=\linewidth]{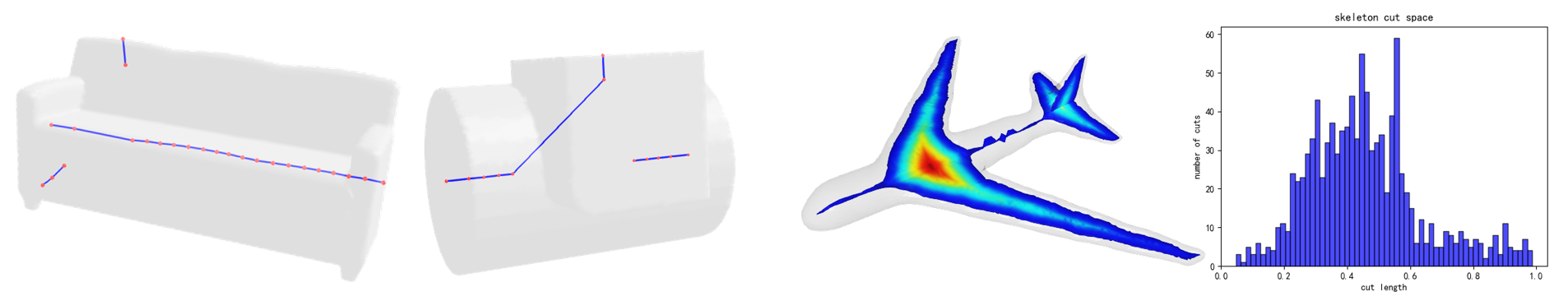}
        \put(60,-6) {\small (a)}
        \put(195, -6) {\small (b)}
    \end{overpic}
    \vspace{-3mm}
    \caption{Analysis of the related skeleton-based methods, i.e., GCD and Skel-Cut. (a) Curve skeleton representations for two models given by the generalized cylinders. (b) The visualization of the metrics used in Skel-Cut: medial geodesic function (MGF) (left) and the cut-length histogram (right).}
    \label{fig:analysis-skeleton}
\end{figure}

In this section, we evaluate our method with comparisons to two representative segmentation methods based on skeletal representations, i.e., generalized cylinder decomposition (GCD) \cite{zhou2015generalized} and skeleton cut space
analysis (Skel-Cut) \cite{feng2015skelcut}. The GCD method, in essence, is still built on the geometrical simplicity given by the 1D-curve representation, but it is defined as generalized cylindricity. Therefore, the method shows its defects when applied to complex geometries and non-cylindrical shapes. As the two shapes shown in Fig.~\ref{fig:analysis-skeleton}~(a), the curve skeletons produced by the generalized cylinders cannot faithfully represent the topology of the original shapes.

A more closely relevant method is Skel-Cut, given that the skeletal representations they use also contain medial surfaces. However, the metric to analyze the cut space is solely based on the medial geodesic function (MGF) \cite{dey2006defining}, of which computation is correlated with the MAT but essentially a distance metric on the surface mesh. Since the Skel-Cut method does not exploit the intrinsic properties of the MAT such as the radius, branches and topology, it is more tailored to the shapes that have distinguishable distributions of the MGF-based cut-length, such as organic shapes. Fig.\ref{fig:analysis-skeleton}~(b) visualizes the MGF of a shape and the corresponding cut-length histogram; it is very difficult to find peaks and valleys on the histogram to differentiate multiple parts based on the algorithm of Skel-Cut.

\begin{figure}[!htb]
\centering
    \begin{overpic}[width=\linewidth]{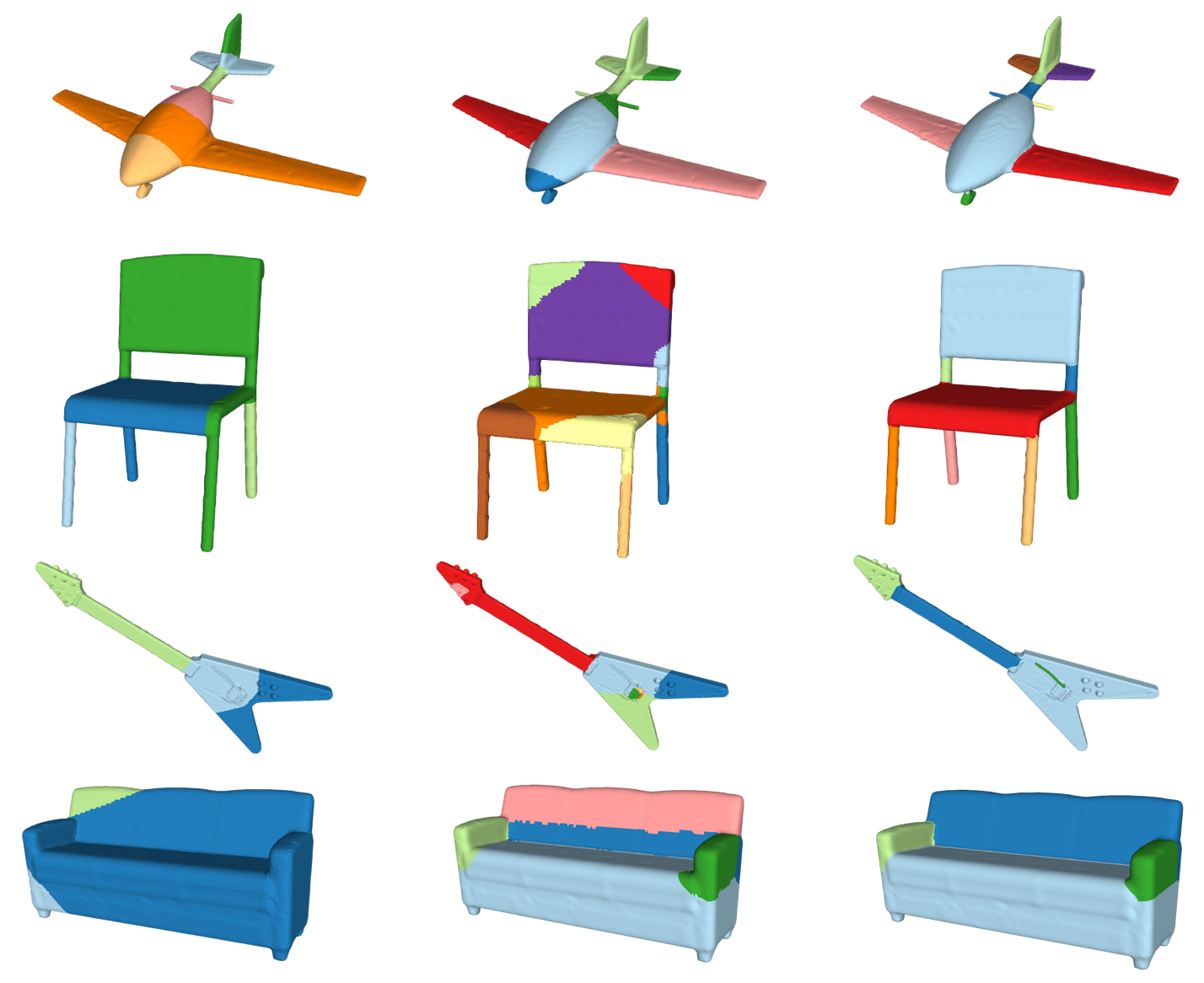}
        \put(30,-6) {\small GCD}
        \put(110, -6) {\small Skel-Cut}
        \put(195, -6) {\small SEG-MAT}
    \end{overpic}
    \vspace{-3mm}
    \caption{Qualitative comparison with representative skeleton-based methods, i.e., Generalized cylinder decomposition \cite{zhou2015generalized} (GCD) and Skel-Cut \cite{feng2015skelcut}.}
    \label{fig:cmp-skeleton}
\end{figure}

\begin{table}[!htb]
\centering
\caption{Quantitative comparison with the skeleton-based methods on the PSB. The CE is reported as the mean of GCE and LCE; the HD is report as the mean of HD, HD-Rm and HD-Rf. The metrics are scaled by 1000 for better readability.}
\setlength{\tabcolsep}{3.5mm}
\begin{tabular}{c|cccc}
\hline
         & RI             & CD             & CE            & HD             \\\hline
GCD      & 159.4          & 252.2          & \textbf{64.7} & 125.3          \\
Skel-Cut & 200.3          & 288.5          & 124.6         & 173.9          \\
SEG-MAT  & \textbf{114.6} & \textbf{212.1} & 79.6          & \textbf{117.1} \\\hline
\end{tabular}
\label{tab:cmp-skeleton}
\end{table}

We compare our method with these two methods using the PSB dataset and a set of shapes from the ShapeNet. The qualitative results are shown in Fig. \ref{fig:cmp-skeleton} and the quantitative results in Table. \ref{tab:cmp-skeleton}. The comparisons demonstrate the generality and robustness of our method to complex geometries especially when a shape is composed of many components with varying properties. For the time efficiency, our method is significantly faster than the others; the average computation time for our method is less than 10 seconds, while the GCD is 134 seconds and Skel-Cut 97 seconds. Also, note we use consistent parameter settings (the parameters of GCD need to be tuned on different categories).

\subsection{Ablation studies}
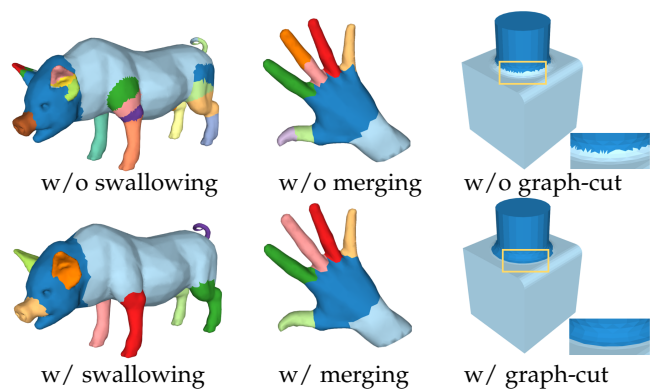
\begin{figure}[!htb]
\centering
    \begin{overpic}[width=\linewidth,]{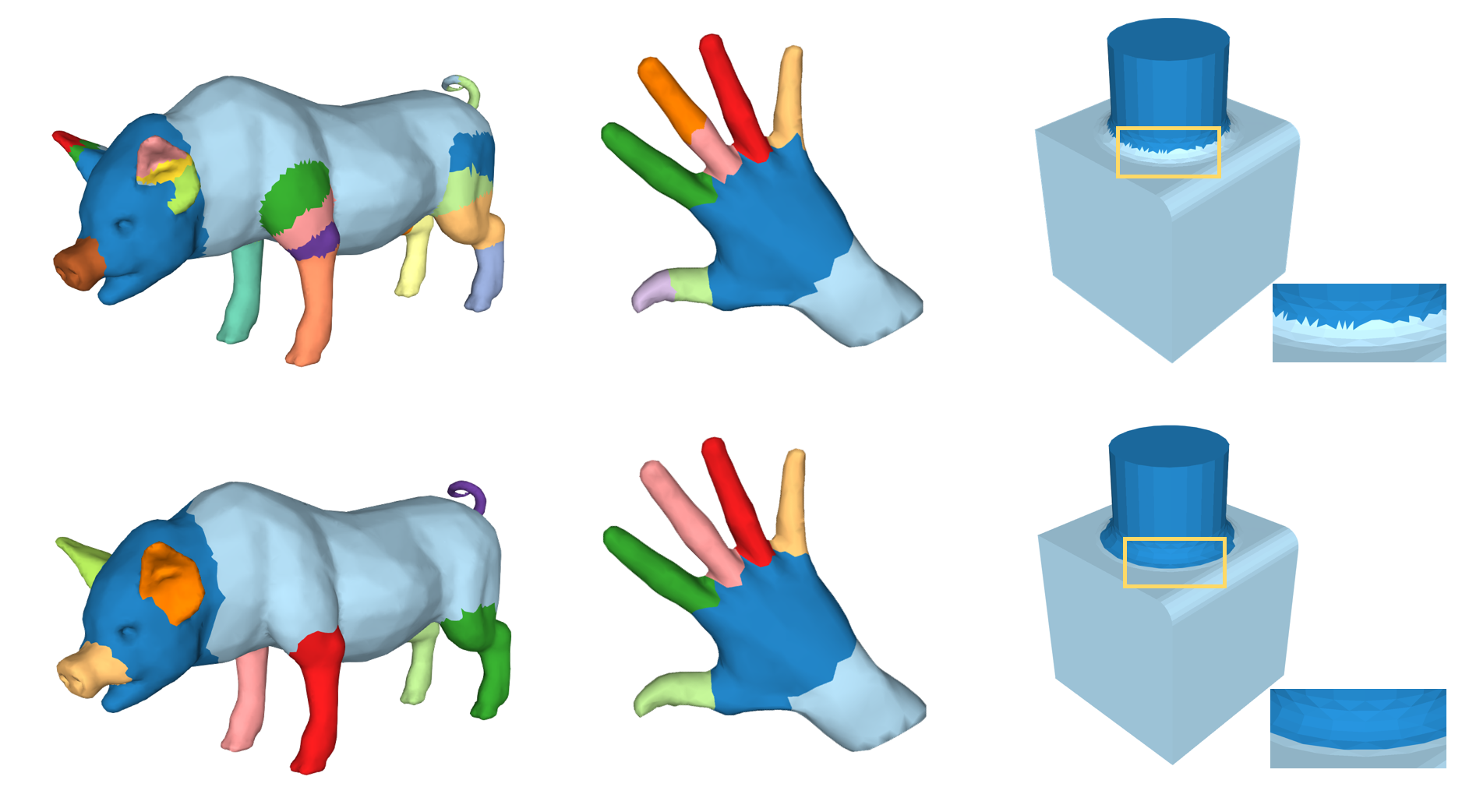}
        \put(20, 70) {\small w/o swallowing}
        \put(20, -3) {\small w/ swallowing}
        \put(110, 70) {\small w/o merging}
        \put(110, -3) {\small w/ merging}
        \put(180, 70) {\small w/o graph-cut}
        \put(180, -3) {\small w/ graph-cut}
    \end{overpic}
    \vspace{-4mm}
\caption{Ablation study on some modules of our algorithm: swallowing, merging and graph-cut.}
\label{fig:ablation}
\end{figure}

We verify the necessity of the key modules in our pipeline by conducting ablation studies. We alternatively remove each module, i.e., swallowing, merging and graph-cut, and evaluate the performance. Fig. \ref{fig:ablation} shows qualitative segmentation results using different configurations, from which we can observe that: (1) the swallowing process effectively helps remove fragmented patches caused by the unstable ``spikes''; (2) the merging process combines the similar parts caused by local variation, giving simpler and more integral segmentation; (3) the graph-cut leads to smoother cut boundaries.  We also report the quantitative evaluation results on the whole dataset in Table. \ref{tab:ablation}.  

\begin{table}[!htb]
\caption{Quantitative ablation study on the different modules in the SEG-MAT pipeline. We report the Rand Index scores on the PSB dataset for different configurations.}
\centering
\setlength{\tabcolsep}{1.5mm}
\begin{tabular}{c|cccc}\hline
          & w/o        & w/o     & w/o       & full           \\
          & swallowing & merging & graph-cut & configuration  \\\hline
RandIndex & 0.135      & 0.129   & 0.127     & \textbf{0.114} \\\hline
\end{tabular}
\label{tab:ablation}
\end{table}

\subsection{Robustness against noise}
\label{sec:anti-noise}
\begin{figure}[t]
    \begin{overpic}[width=\linewidth]{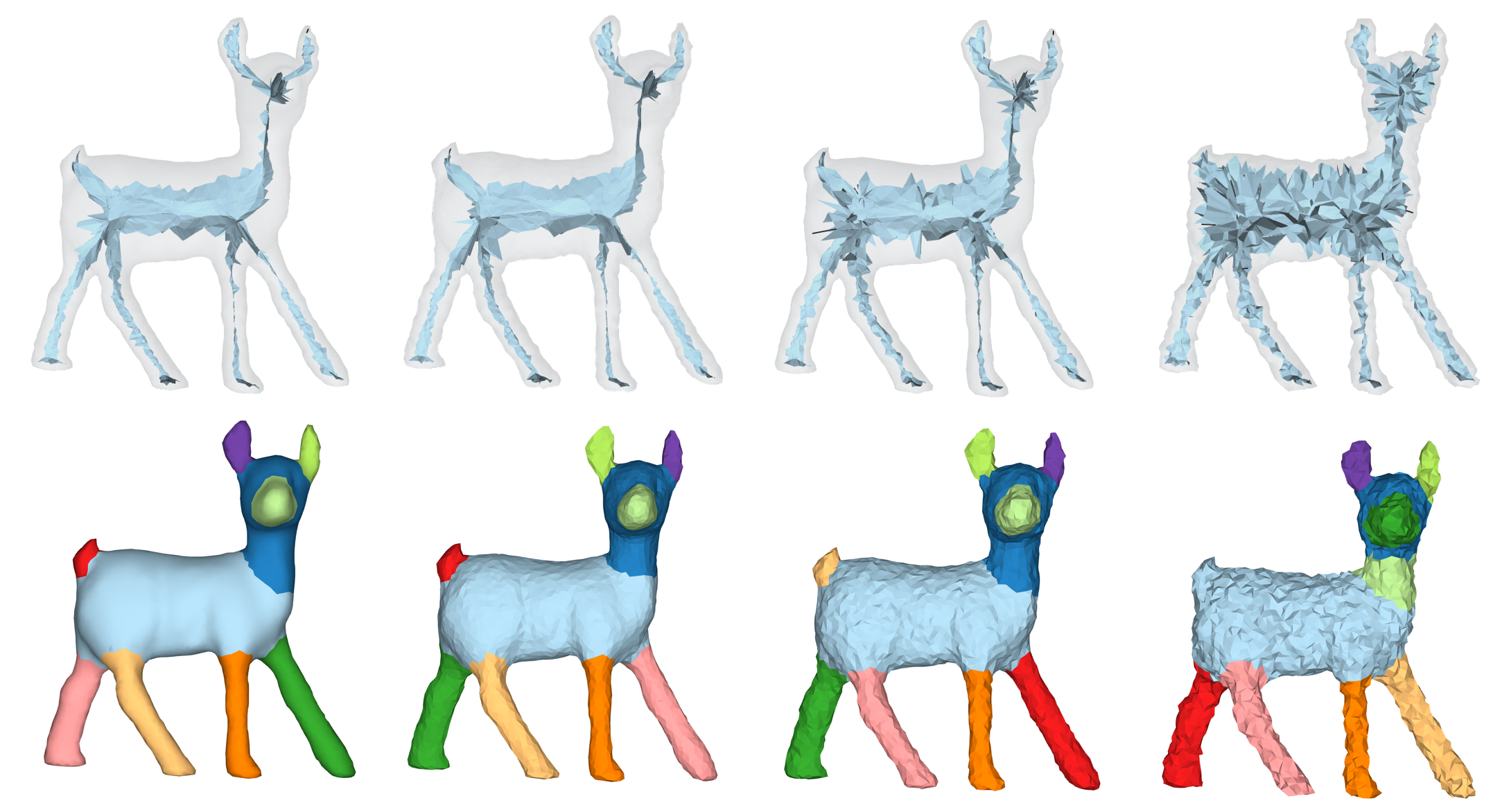}
        \put(13, -7) {\small $\sigma=0.0\%$}
        \put(75, -7) {\small $\sigma=0.2\%$}
        \put(137, -7) {\small $\sigma=0.5\%$}
        \put(200, -7) {\small $\sigma=1.0\%$}
    \end{overpic}
    \vspace{0mm}
    \caption{Robustness of SEG-MAT against surface noise. The first row shows the medial meshes and the second row shows the corresponding segmentation results. We add to each mesh vertex a white Gaussian noise with the mean value equal to zero and the standard deviation $\sigma=0.0\%, 0.2\%, 0.5\%, 1.0\%$ of the length of the diagonal of bounding box respectively. Using a same set of parameter values, SEG-MAT is able to produce almost consistent segmentation results.}
    \label{fig:noise}
\end{figure}

\begin{figure}[!htb]
\centering
    \includegraphics[width=\linewidth]{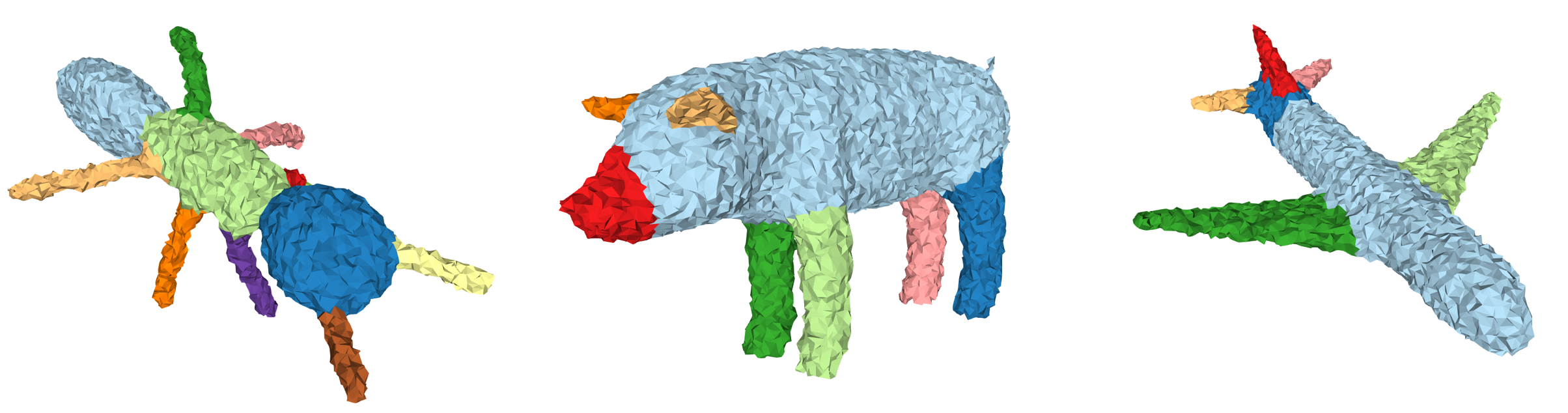}
    \vspace{-7mm}
    \caption{More segmentation results of shapes with significant noise, while SEG-MAT is still able to segment these shapes into meaningful parts. }
    \label{fig:morenoise}
\end{figure}

Although the MAT itself is very sensitive to boundary noise, our algorithm is resilient in the presence of significant noise. Fig. \ref{fig:noise} shows an example of applying our algorithm to a model with increasing noise, with all the parameters being fixed. It can be observed that SEG-MAT produces almost consistent segmentation results at different noise levels. The robustness of SEG-MAT to surface noise is due to the use of the region growing technique and the swallowing process. Region growing is a greedy strategy, and hence we can generate stable regions by greedily finding similar neighbors rather than globally considering the insignificant branches. More importantly, the ``swallowing'' (Sec. \ref{sec:swallowing}) process merges the noisy branches into the stable regions, and thus the noisy branches will not be considered as valid parts. Therefore, the algorithm only focuses on the backbone of the MAT and is insensitive to noise. More results are shown in Fig. \ref{fig:morenoise} for some shapes with significant noise.

\subsection{Parameter analysis}
\label{subsec:parameterAnalysis}
Since our method is geometry-driven, like all the existing methods, there is a set of parameters in the algorithm to adapt to the geometrical information from different perspectives. We use consistent parameter settings for producing all the results for comparison, which demonstrates that our default values work well for general models, such as these shapes in the PSB~\cite{chen2009benchmark} and ShapeNet~\cite{chang2015shapenet}. 

\begin{figure}[!htb]
\centering
    \begin{overpic}[width=\linewidth]{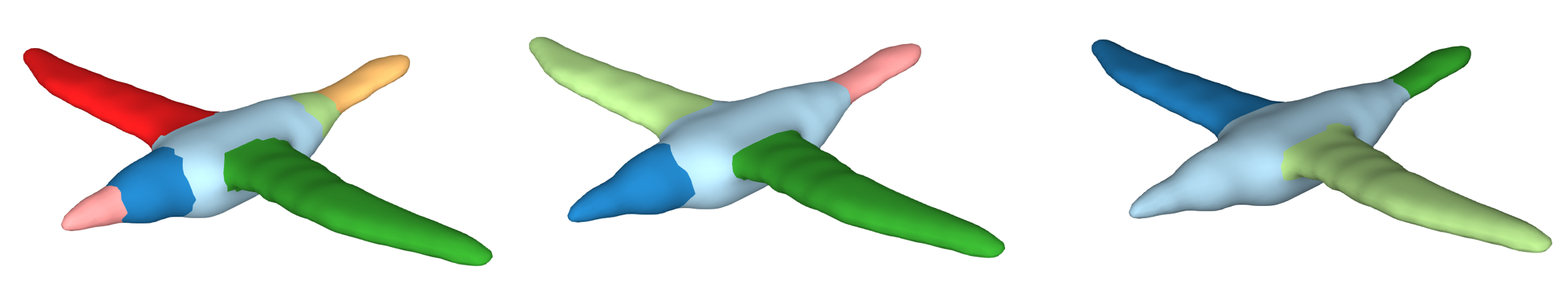}
        \put(20, -6) {\small $\delta=0.01$}
        \put(103, -6) {\small $\delta=0.015$}
        \put(193, -6) {\small $\delta=0.03$}
    \end{overpic}
    \caption{Segmentation results for a bird model with an increasing growing threshold $\delta$. A smaller $\delta$ produces more fine-grained segmentation that captures more detail.}
    \label{fig:growpara}
\end{figure} 

\begin{figure}[!htb]
    \vspace{-2mm}
    \begin{overpic}[width=\linewidth]{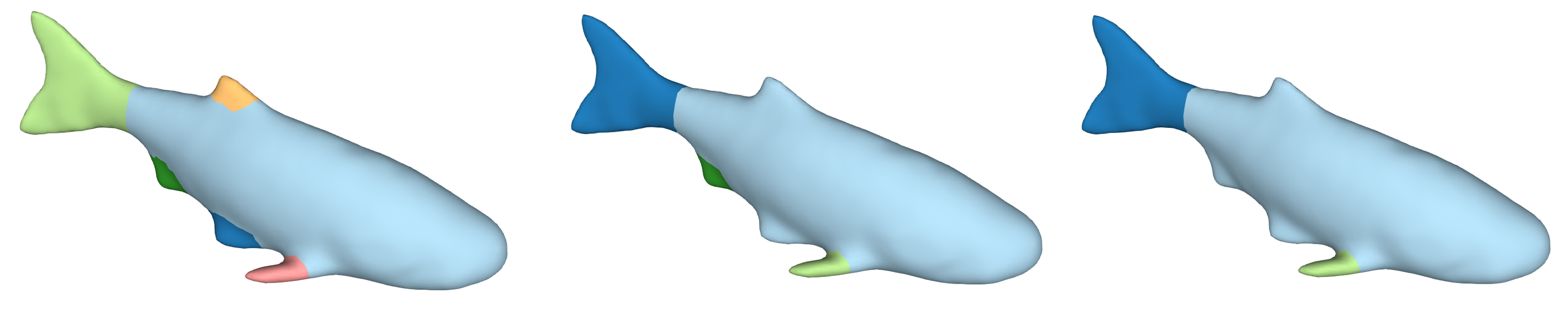}
        \put(20, -6) {\small $\eta=0.2\%$}
        \put(108, -6) {\small $\eta=0.3\%$}
        \put(193, -6) {\small $\eta=0.4\%$}
    \end{overpic}
    \vspace{-2mm}
    \caption{Segmentation results for a fish model with increasing values of the minimal region threshold $\eta$. A larger $\eta$ results in removing tiny parts in the segmentation result.}
    \label{fig:minregion}
\end{figure}

\begin{table}[!htb]
\caption{Evaluation on the sensitivity of the other parameters. We report the maximal and minimal Rand Index score achieved within the range of parameter test.}
\centering
\begin{tabular}{c|cccc}\hline
Parameter   & Function             & Default value & RI-min & RI-max \\\hline
$\alpha$  & weight of Eq. \ref{eq:ma_cost} & 0.05    & 0.114  & 0.118  \\
$\lambda$ & weight of Eq. \ref{eq:growingFull} & 1.5     & 0.113  & 0.121  \\
$\omega$  & weight of Eq. \ref{eq:graphcut} & 0.3     & 0.114  & 0.119 \\\hline
\end{tabular}
\label{tab:para_sensitivity}
\end{table}

It is however possible for users to tune some parameters to adjust the desired segmentation. We only leave two necessary parameters of region growing algorithm, i.e., the growing threshold $\delta$ and the minimal region threshold $\eta$ (see Sec.~\ref{sec:geo_dec}), to users to generate different levels of details; other parameters are recommended to be fixed to the default values obtained by our extensive experiments. We discuss the effect of different values of these two parameters on the segmentation results for a better understanding of the mechanism of SEG-MAT. 

The growing cost tolerance $\delta$ affects the sensitivity of the segmentation to the geometrical variation. As shown in Fig.~\ref{fig:growpara}, a smaller $\delta$ captures more detailed geometrical change and results in a finer-grained segmentation. We use $\delta=0.015$ for the evaluations in this paper.

The minimal region threshold $\eta$ is used to control the minimum size of a part, while using a small $\eta$ leads to preserving the tiny parts of a 3D shape. See Fig. \ref{fig:minregion} for the validation experiments using different values of $\eta$. The default value is $\eta=0.2\%$ for the evaluations in this paper.

For the other parameters, following \cite{wcseg}, we change their values in a range within $30\%$ ($\pm 15\%$ added to each value with $5\%$ changed each step). Accordingly, we report the range of the variation of the Rand Index on the PSB dataset, which is shown in Table. \ref{tab:para_sensitivity}. It can be seen that the segmentation quality is largely similar.

\subsection{More applications}
\label{sec:more_applications}
We extend our method to make it applicable to more tasks. In this section, we demonstrate two applications, i.e., primitive-based abstraction and point cloud segmentation.
\para{Primitive-based abstraction}
There has been growing interest in parsing a 3D shape into a primitive-based representation \cite{abstractionTulsiani17, zhu_siga18, sun2019hierarchy}, of which goal is to assemble a target shape using a set of volumetric primitives. A good primitive-based approximation is geometrically expressive and compact, which means it can abstract complex 3D shape structures using the fewest possible primitives. To this end, given a 3D shape, we first run our algorithm to segment it into a set of structurally meaningful parts, and then find a minimal oriented bounding box (MOBB) \cite{mobb2001} for each part. The results are shown in Fig.~\ref{fig:prim_results}.

\begin{figure}[!htb]
\centering
    \includegraphics[width=\linewidth]{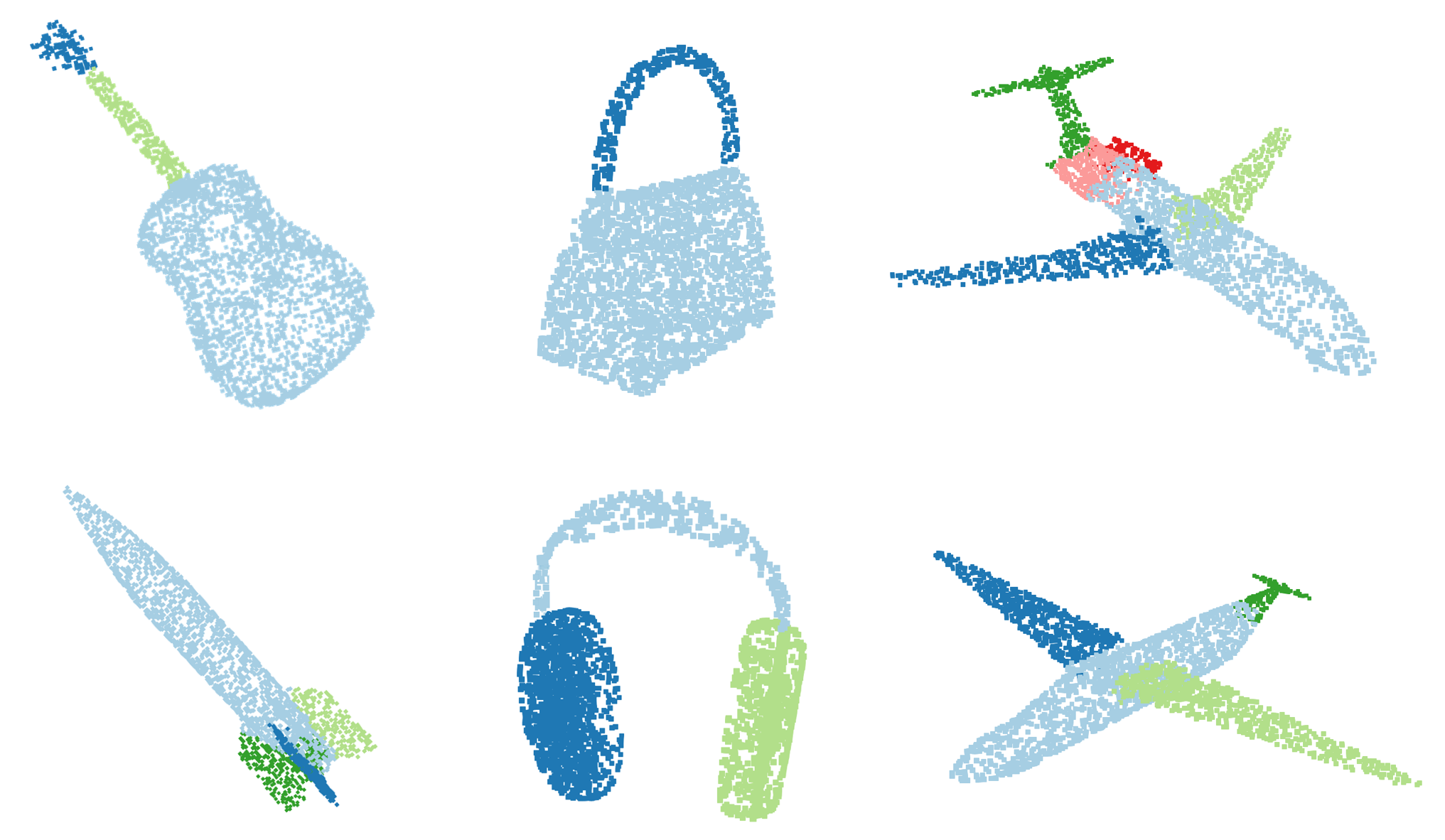}
    \caption{Segmentation results of SEG-MAT for point clouds. }
    \label{fig:pcresults}
\end{figure}

\para{Point cloud segmentation}
SEG-MAT can also be used for the segmentation of point clouds. We first use Deep Point Consolidation \cite{wu2015deep} to compute a meso-skeleton from a noisy or incomplete input, and then use the closest distance between each medial point to the input point cloud as the approximation of the radius. Then the MAT graph is constructed and the normals are estimated using the k-nearest neighbors of each medial point. Since there are no surfaces on the meso-skeleton, we simply remove the medial primitive term $C_{mp}$ in Eq.~\ref{eq:growingFull}. Some results are shown in Fig.~\ref{fig:pcresults}.

\section{Conclusion}
We present a simple, robust and efficient method, called {\em SEG-MAT}, for 3D shape segmentation using the medial axis transform (MAT). The use of the structural and geometrical information encoded in the MAT makes SEG-MAT effective for segmenting arbitrary shapes across a wide range of complexities and noise levels. Given a 3D shape with its MAT, we first perform structural decomposition by detecting the joints using the simplified MAT, and then perform geometrical decomposition by region growing. Finally, the segmentation results are transferred from the MAT to the shape surface. The extensive evaluations and comparisons with the existing segmentation methods demonstrate that SEG-MAT is a superior and competitive geometry-driven method for real-world applications of 3D shape analysis. 

\begin{figure}[!htb]
    \centering
    \begin{overpic}[width=0.95\linewidth]{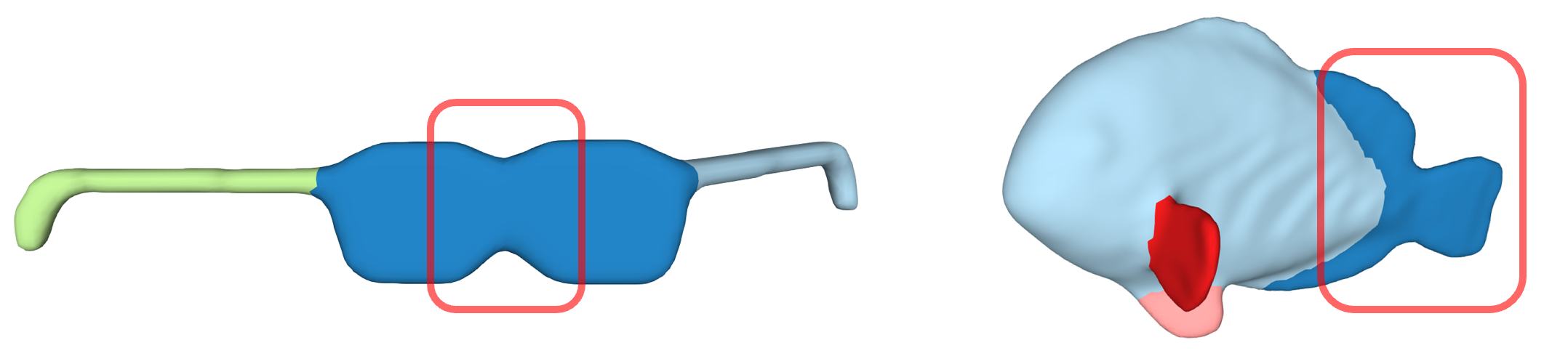}
        \put(63, -8) {\small (a)}
        \put(190, -8) {\small (b)}
    \end{overpic}
    \caption{Some failure cases of SEG-MAT. (a) Two pieces of the glasses should be segmented into two distinct parts; (b) the tail and the fin of the fish should be decomposed.}
    \label{fig:limitation}
\end{figure}

\para{Limitations and future work}  SEG-MAT would fail to segment the parts that have flat geometries and constant thickness. Fig. \ref{fig:limitation} shows two examples, each of which contains a part that is not segmented by SEG-MAT. Here, the MAT surface of such a part is a single flat sheet and the radius function is constant. SEG-MAT fails to segment it because there is neither a large bending angle, nor a radius variation, nor a structural change on the MAT to trigger a segmentation. A potential solution to this problem could be based on the analysis of the medial curve of the MAT surface sheet to detect the narrow passage connecting these flat parts, following the idea of the erosion function of the MAT \cite{yan2016erosion}.

\ifCLASSOPTIONcompsoc
  \section*{Acknowledgments}
 We would like to thank the anonymous reviewers for their valuable feedback and Yiling Pan for her help with data processing. This work is supported by the Gottfried Wilhelm Leibniz program by DFG and the grants from National Natural Science Foundation of China (NSFC) (No. 61772301 and No. 61772016).
 
\else
  \section*{Acknowledgment}
\fi


\ifCLASSOPTIONcaptionsoff
  \newpage
\fi




\bibliographystyle{IEEEtran}
\bibliography{references}




 \end{document}